\documentclass[10pt,twocolumn,prd,nofootinbib]{revtex4}

\usepackage{graphicx}
\usepackage{psfrag}
\usepackage{amsmath}
\newcommand{\beq}{\begin{equation}}
\newcommand{\eeq}{\end{equation}}
\def\beqa{\begin{eqnarray}}
\def\eeqa{\end{eqnarray}}
\def\p{\partial}

\def\lap{\lower.5ex\hbox{$\; \buildrel < \over \sim \;$}}
\def\gap{\lower.5ex\hbox{$\; \buildrel > \over \sim \;$}}
\def\lesssim{\lap}

\begin{document}

\title{Probabilities in the inflationary multiverse}
\author{Jaume Garriga$^1$, Delia Schwartz-Perlov$^2$, Alexander
Vilenkin$^2$, and Sergei Winitzki$^3$} \affiliation{
$^1$ Departament de F{\'\i}sica Fonamental, Universitat de Barcelona,\\
Mart{\'\i}\ i Franqu{\`e}s 1, 08193 Barcelona, Spain\\
$^2$ Institute of Cosmology, Department of Physics and Astronomy,\\
Tufts University, Medford, MA 02155, USA,\\
$^3$ Department of Physics, Ludwig-Maximilians University,\\
Theresienstr.~37, 80333 Munich, Germany
}

\begin{abstract}

Inflationary cosmology leads to the picture of a ``multiverse,"
involving an infinite number of (spatially infinite)
post-inflationary thermalized regions, called pocket universes. In
the context of theories with many vacua, such as the landscape of
string theory, the effective constants of Nature are randomized by
quantum processes during inflation. We discuss an analytic
estimate for the volume distribution of the constants within each
pocket universe. This is based on the conjecture that the field
distribution is approximately ergodic in the diffusion regime,
when the dynamics of the fields is dominated by quantum
fluctuations (rather than by the classical drift). We then propose
a method for determining the relative abundances of different
types of pocket universes. Both ingredients are combined into
an expression for the distribution of the constants in pocket
universes of all types.

\end{abstract}
\maketitle

\section{Introduction}

The fundamental theory of nature may admit multiple vacua with different
low-energy constants. If there were just a few vacua, as in standard GUT models,
then a few observations would determine which one corresponds to the real world.
Predictions would then follow for every other observable in the low energy
theory. However, it has recently been realized that in the context of string
theory there may be a vast landscape of possibilities, with googols of vacua to
scan \cite{duff,Susskind,Bousso}. Many of these may look very much like our own
vacuum, except for slight variations in the values of the constants. On the
surface, this seems to undermine our ability for predicting these values, even
after a systematic examination of the landscape.

Cosmology, on the other hand, suggests that rather than giving up
on our ability to make predictions, we may in fact broaden their
scope. Thus, instead of trying to determine from observations
which vacuum is ours, we may try to determine, from the theory,
what is the probability for the observation of certain values of
the constants.\footnote{Note that this question is relevant even
if the number of vacua is small.} Indeed, eternal inflation
\cite{AV83,Linde86} leads to the picture of a ``multiverse,''
where constants of nature take different values in different
post-inflationary regions of spacetime. Observers bloom in such
thermalized regions at places where the conditions are favorable,
much like wildflowers at certain spots in the forest. Given a
reference class of observers, we can ask what is the probability
distribution for the values of the constants that they will
measure. This approach was suggested in \cite{alex} and further
developed in \cite{GTV98,AV98,VVW,GV01}. It leads to the following
formal expression for the probability of observations, \beq
P_{obs}(X) \propto P(X)\ n_{obs}(X).\label{pobs} \eeq $P(X)$ is
the volume fraction\footnote{$P(X)$ is sometimes referred to as
the ``prior distribution." This name is motivated by Carter's
original discussion of anthropic selection in terms of Bayes'
rule. To avoid confusion with the usage of prior distributions in
other contexts, here we shall simply call it the thermalized
volume distribution. Note that the concept of a prior distribution
for a set of parameters is also used when fitting observational
data to a given model. There, the ``prior" is often no more than a
guess, representing our ignorance or prejudice about, say, the
allowed range of the parameters. On the contrary, $P(X)$ is here a
quantity which should be calculable from the theory.} of
thermalized regions with given values of the constants $X$, and
$n_{obs}(X)$ is the number of observers in such regions per unit
thermalized volume.\footnote{$n_{obs}(X)$ is often referred to as
the anthropic factor. Readers who dislike anthropic arguments are
encouraged to think in terms of reference classes of their own
choice.}

Even though the dynamics responsible for the randomization of the
constants during inflation is well understood, the calculation of
probabilities has proven to be a rather challenging problem. The root
of the difficulty is that the volume of thermalized regions with any
given values of the constants is infinite (even for a region of a
finite comoving size). To compare such infinite volumes, one has to
introduce some sort of a cutoff. For example, one could include only
regions that thermalized prior to some time $t_{c}$ and evaluate
volume ratios in the limit $t_{c}\rightarrow\infty$. However, one
finds that the results are highly sensitive to the choice of the
cutoff procedure (in the example above, to the choice of the time coordinate
$t$~\cite{LLM94,LLM+}; see also \cite{Guth,Tegmark} for a recent discussion.)
The reason for the cutoff dependence of probabilities is that the volume
of an eternally inflating universe is growing exponentially
with time. The volumes of regions with all possible values of the
constants are growing exponentially as well. At any time, a
substantial part of the total thermalized volume is ``new'' and thus
close to the cutoff surface. It is not surprising, therefore, that the
result depends on how that surface is drawn.

\begin{figure}
\begin{center}
\psfrag{t}{$t$}\psfrag{x}{$x$}
\includegraphics[width=3in]{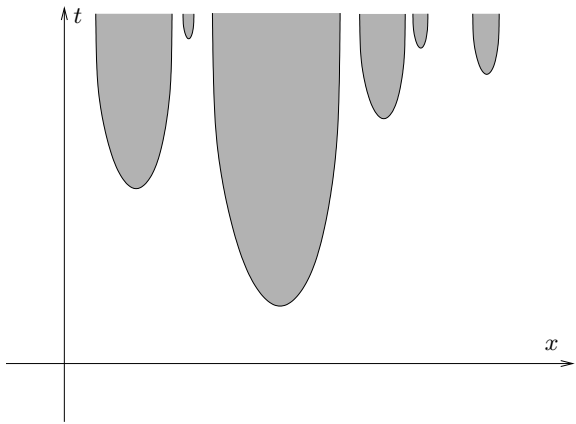}

\caption{Structure of eternally inflating spacetime. Thermalized
regions are shown by dark shading.} \label{shades}
\end{center}
\end{figure}

As suggested in \cite{GTV98,AV98}, this difficulty can be circumvented
by switching from a global distribution, defined with the aid of
some global time coordinate, to a distribution based on individual
thermalized regions.  The spacetime structure of an eternally
inflating universe is illustrated in Fig.~\ref{shades}. The
vertical axis is the proper time measured by comoving observers,
and the horizontal axis is the comoving coordinate. Thermalized
regions are marked by grey shading. Horizontal slices through this
spacetime give ``snapshots'' of a comoving volume at different
moments of (global) time. Initially, the whole volume is in the
inflating state. While the volume expands exponentially, new
thermalized regions are constantly being formed. These regions
expand into the inflating background, but the gaps between them
also expand, making room for more thermalized regions to form. The
thermalization surfaces at the boundaries between inflating and
thermalized spacetime regions are 3-dimensional, infinite,
spacelike hypersurfaces. The spacetime geometry of an individual
thermalized region is most naturally described by choosing the
corresponding thermalization surface as the origin of time. The
thermalized region then appears as a self-contained infinite
universe, with the thermalization surface playing the role of the
big bang. Following Alan Guth, we shall call such infinite domains
``pocket universes.'' All pocket universes are spacelike-separated
and thus causally disconnected from one another. In models where
false vacuum decays through bubble nucleation, the role of pocket
universes is played by individual bubbles.

The proposal of Refs.~\cite{GTV98,AV98} applies to the special case
when all pocket universes are statistically equivalent. This happens,
for instance, if there is only one low-energy vacuum, and the
variation of the constants $X$ is due to long-wavelength fluctuations
of some nearly massless fields. In this case, one may use any one of
these pockets in order to calculate $P(X)$, which is then proportional
to the volume fraction occupied by the corresponding regions in the
pocket universe. This fraction can be found by first evaluating it
within a sphere of large radius $R$ and then taking the limit
$R\to\infty$.

Although the above proposal is conceptually satisfactory, it remains
unclear how it should be implemented in practice. The most direct
method, a numerical simulation, runs into severe computational
limitations \cite{LLM94,VVW}, and analytic methods have been developed
only for very special cases. Moreover, if two or more vacua are
mutually separated by inflating domain walls, then they cannot coexist
in the same pocket universe. In this situation (which is expected to
be quite generic in the landscape of string theory) there are distinct
types of pocket universes, and we have to face the problem of
comparing probabilities of different pockets.  The purpose of the
present paper is to try to improve on the approach of
Refs. \cite{GTV98,AV98,VVW} by addressing both of these issues.

In Section II we discuss an analytic estimate for the volume
distribution of the constants within each pocket, based on the
conjecture that the field distribution is approximately ergodic in
the diffusion regime. When there are different types of pockets,
the same approach can be used in order to find the internal volume
distribution $P(X;j)$ in a pocket of type $j$. Then, in Section
III, we introduce a new object $p_j$ which characterizes the
relative abundance of each type of pocket universe. The
aficionados may be aware of previous exploratory definitions of
$p_j$, which were nevertheless afflicted with certain drawbacks
\cite{GV01}. For comparison, in Section IV we discuss one of these
alternatives and in Section V we illustrate both definitions with
some examples. Section VI discusses how the two objects $P(X;j)$
and $p_j$ may be combined in order to find the full distribution
for the constants. Our conclusions are summarized in Section VII.

\section{Probabilities within a pocket universe}

\subsection{Symmetric inflaton potential}

\begin{figure}
\begin{center}
\includegraphics[width=4in]{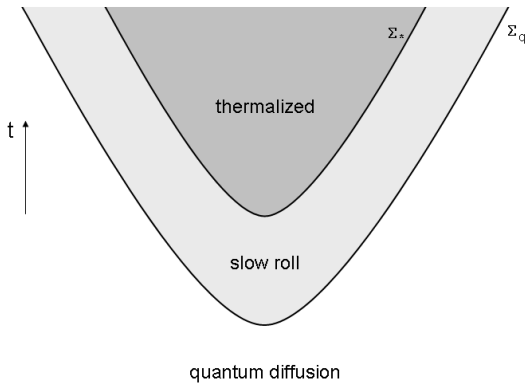}

\caption{Structure of a pocket universe.} \label{pocket}
\end{center}
\end{figure}

The spacetime structure of a pocket universe is illustrated in
Fig.~\ref{pocket}. The surface $\Sigma_*$ in the figure is the
thermalization surface. It is the boundary between inflating and
thermalized domains of spacetime, which marks the end of inflation
and plays the role of the big bang in the pocket universe. The
surface $\Sigma_q$ is the boundary between the stochastic domain,
where the dynamics of the inflaton field is dominated by quantum
diffusion, and the deterministic domain, where the dynamics is
dominated by the deterministic slow roll. Thus, $\Sigma_q$ marks
the onset of the slow roll.

The surface $\Sigma_*$ is always spacelike. We shall assume that
$\Sigma_q$ is spacelike as well. This can be arranged if
$\Sigma_q$ is chosen not right at the diffusion boundary, but
somewhat into the slow roll domain. We shall comment on the
precise definition of $\Sigma_q$ at the end of Section II.C.

The probability $P(X)dX$ for a variable $X$ is defined as the
volume fraction on the thermalization hypersurface $\Sigma_*$ with
given values of $X$. It can be expressed as \beq P(X)\propto
P_q(X)Z^3(X), \label{Pq} \eeq where $P_q(X)$ is the distribution
(volume fraction) of $X$ on $\Sigma_q$, and $Z(X)$ is the volume
expansion factor during the slow roll in regions with the value
$X$. For simplicity, we shall identify the variable $X$ with the
scalar field responsible for its value. Here, we have assumed that
the diffusion of X can be neglected during the slow roll. We shall
assume also that $X$ interacts very weakly with the inflaton, so
that it does not change appreciably during the slow roll.
Otherwise, we would have to include an additional Jacobian factor
\beq \det\left[{\partial X(t_q)/\partial X(t_{obs})}\right],
\label{jacobian} \eeq in Eq.~(\ref{Pq}) \cite{AV98}. Here, $t_q$
and $t_{obs}$ indicate, respectively, the time when slow roll
begins and the time when observations are made.

The expansion factor $Z(X)$ can be found from \beq Z(X)\approx
\exp\left[4\pi\int_{\phi_{*X}}^{\phi_{qX}} \frac{H(\phi,X)}
{H'(\phi,X)} d\phi\right], \label{Z} \eeq where $\phi$ is the
inflaton field, \beq H(\phi;X)=[8\pi V(\phi,X)/3]^{1/2} \label{HV}
\eeq is the inflationary expansion rate, $V(\phi,X)$ is the
inflaton potential, the prime stands for a derivative with respect
to $\phi$, and we use Planck units throughout this paper. We
denoted by $\phi_{qX}$ and $\phi_{*X}$ the values of $\phi$ at the
boundary surfaces $\Sigma_q$ and $\Sigma_*$, respectively. These values
are defined by the conditions \beq H'/H^2(\phi_{qX},X)\sim 1
\label{phiq} \eeq and \beq H'/H(\phi_{*X},X)\sim 1. \label{phi*}
\eeq The subscript $X$ in $\phi_{qX}$ indicates that the value of
$\phi$ where slow roll begins is generally $X$-dependent, $\phi =
\phi_q(X) \equiv \phi_{qX}$, and similarly for $\phi_{*X}$.

The distribution $P_q(X)$ can in principle be determined from
numerical simulations of the quantum diffusion regime. Some useful
techniques for this type of simulation have been developed in
\cite{LLM94,VVW,VW}. However, the simulations quickly run into
computational limits, due to the exponential character of the
expansion. Analytic techniques for the calculation of $P(X)$ are therefore very desirable.

A special case where the analysis is trivial is the class of
models where the potential is independent of $X$ in the diffusion
regime. Then $P_q(X)$ can be determined from symmetry, \beq P_q(X)
= {\rm const}. \label{Pqflat} \eeq An example is a ``new''
inflation type model with a complex inflaton, $\phi = |\phi| \exp
(iX)$. Inflation occurs near the maximum of the potential at
$\phi=0$, and we assume in addition that the potential is
symmetric near the top, $V = V(|\phi|)$. Equation~(\ref{Pqflat})
follows if this property holds throughout the diffusion regime. In
such models, the distribution (\ref{Pq}) reduces to \cite{AV98}
\beq P(X)\propto Z^3(X). \label{PZ} \eeq Thus, the volume
distribution is simply determined by the expansion factor during
the slow roll.

In models where the inflaton potential does not have the assumed
symmetry, the factor $P_q(X)$ may provide an additional dependence on
$X$. We now turn to the discussion of this more general case.

\subsection{Eternal inflation without thermalization}

Let us first consider a model of eternal inflation without any
low-energy vacua, where the dynamics is dominated by diffusion
everywhere in the potential. The stochastic evolution of fields in
the course of eternal inflation is described by the distribution
function \beq F(\phi_a,t), \label{F} \eeq where $\phi_a$ stands
collectively for the inflaton and other light fields. This
distribution satisfies the Fokker-Planck (FP) equation
\cite{AV83,Starobinsky} \beq {\p F\over{\p t}}=-{\p
J_a\over{\p\phi_a}}, \label{FP} \eeq where the flux $J_a(\phi,t)$
is given by \beq J_a=-{\p\over{\p\phi_a}}[D(\phi)F] +v_a(\phi)F.
\label{J} \eeq Here, \beq D(\phi)=H(\phi)^{\beta+2}/8\pi^2
\label{D} \eeq is the diffusion coefficient, \beq
v_a(\phi)=-{1\over{4\pi}}H(\phi)^{\beta-1}{\p\over{\p\phi_a}}
H(\phi) \label{v} \eeq is the ``drift'' velocity of the slow roll,
and as before $H(\phi)=[8\pi V(\phi)/3]^{1/2}$ is the inflationary
expansion rate.

The parameter $\beta$ in Eqs.~(\ref{D}) and (\ref{v}) represents the
freedom of choosing the time variable $t$, which is assumed to be
related to the proper time of comoving observers $\tau$ by
\beq
dt=[H(\phi(\tau))]^{1-\beta}d\tau,
\label{t}
\eeq
where the functions $\phi(\tau)$ are taken along the observer's world
lines. Hence, $\beta=1$ corresponds to the proper time parametrization
$t=\tau$ and $\beta=0$ corresponds to using the logarithm of the scale
factor as the time variable. (We shall refer to the latter choice as
the scale factor time.)

The FP equation is usually supplemented by boundary conditions at
the thermalization boundary in $\phi$-space, $S_*$, where the
conditions of slow roll are violated, and in some cases at the
Planck boundary $S_p$, where $V(\phi)\sim 1$, and quantum gravity
effects become important. Here, we shall assume that neither of
these boundaries is present. For example, the $\phi$-field space
could be compact, with the potential $V(\phi)$ satisfying
$V(\phi)\ll 1$ and $H'/H^2 \ll 1$ everywhere in this space. Then
the fields $\phi_a$ will drift constantly from one value to
another. The corresponding distribution has a stationary form with
a vanishing flux, \beq J_a = 0. \label{J=0} \eeq

With $J_a$ from (\ref{J}), the solution of Eq.~(\ref{J=0}) is easily
found \cite{Starobinsky}. Up to a normalization constant, it is given by
\beq
F(\phi)=H^{-(2+\beta)}(\phi)e^{\pi/H^2(\phi)}.
\label{solution}
\eeq
Let us now use the scale factor time, $\beta = 0$, which
corresponds
to measuring time in units of the Hubble time $H^{-1}$. With
this choice, Eq.~(\ref{solution}) can be rewritten as
\beq
F(\phi)=H^{-2}(\phi)e^{S(\phi)},
\label{FS}
\eeq
where
\beq
S(\phi)=\pi/H^2(\phi)
\label{S}
\eeq
is the Gibbons-Hawking entropy of de Sitter space.

The solution (\ref{FS}) has a simple physical interpretation. The
distribution $F(\phi,t)$ can be thought of as representing observations
by a comoving observer. The observer can see only her own horizon
region ($h$-region), and the value of $\phi$ should be understood as an average
over that region. The quantum state of the $h$-region is
constantly changing due to de Sitter quantum fluctuations, and the
average $\phi$ is also changing as a result. The distribution function
$F(\phi)$ is proportional to the fraction of time spent in quantum
states with a given $\phi$. The number of microscopic
quantum states corresponding to a given value of $\phi$ is $\exp
[S(\phi)]$, and Eq.~(\ref{FS}) indicates that, apart from the
prefactor, the evolution during eternal inflation is {\it ergodic},
with all quantum states explored with an equal weight.

The prefactor $H^{-2}$ suggests that at higher rates of inflationary
expansion, the quantum state of an $h$-region changes at a higher
pace.  The scale-factor time spent in each quantum state is $\delta t
\propto H^{-2}$ (the proper time $\delta\tau \propto H^{-3}$). We
note, however, that the prefactor in the solution (\ref{FS}) depends
on the choice of ordering of the non-commuting factors $\phi$ and
$\p/\p\phi$ in the FP equation. In Eq.~(\ref{J}) above, we assumed the
Ito factor ordering---the choice suggested by the phenomenological
analysis in \cite{AV99}. A full microscopic derivation would require
inclusion of quantum gravitational fluctuations, and until then the
precise form of the prefactor will remain uncertain. We shall mostly
disregard the prefactor in what follows.

The distribution (\ref{FS}) has an alternative interpretation, which
will also be important for our analysis. It follows from the so-called
physical-volume form of the FP equation \cite{GLM87,Sasaki,LLM94},
\beq
{\p {\tilde F}\over{\p t}}=-{\p {\tilde J}_a\over{\p\phi_a}}
+ 3H^\beta {\tilde F}.
\label{FP'}
\eeq
Here, the function ${\tilde F}(\phi,t)$ characterizes the distribution
of physical volume between regions with different values of $\phi$.
More precisely, ${\tilde F}(\phi,t)d^n\phi$ is the fraction of volume
occupied by regions with $\phi_a$ in the intervals $d\phi_a$ on
hypersurfaces of constant $t$. The flux ${\tilde J}_a$ is given by
(\ref{J}) with $F$ replaced by ${\tilde F}$, and the last term in
(\ref{FP'}) accounts for the expansion of physical volume.

With $\beta=0$, solutions of Eq.~(\ref{FP'}) and
those of Eq.~(\ref{FP}) are closely related \cite{LLM94},
\beq
{\tilde F}(\phi,t)=e^{3t}F(\phi,t).
\label{FF'}
\eeq
This applies, in particular, to the solution (\ref{FS}). On surfaces
of constant $t$, the factor $\exp(3t)$ is a constant, and thus the
volume distribution of $\phi$ along these surfaces is the same as the
time distribution along the worldlines of comoving observers.

To appreciate the significance of this result, imagine dividing
the spacetime into $4D$ cells of size $\sim \delta/H$ with
$\delta\ll 1$. This can be done by first foliating the spacetime
by surfaces of constant scale factor time, $t = 0, \delta,
2\delta, ...$. The timelike separation between consecutive
surfaces is of order $\delta/H$. The next step is to divide the
surface $t=0$ into $3D$ cubes and to extend these cubes along
comoving geodesics all the way to the surface $t=\delta$. As a
result, the spacetime layer between the surfaces $t=0$ and
$t=\delta$ is divided into $4D$ cells. This procedure is similar
to laying bricks, starting with the surface $t=0$, except that the
bricks are somewhat irregular, with their tops being slightly
bigger than their bottoms and their size depending on the local
value of $H$.

Now, we wish to know how $\phi$ is distributed among different
cells. It follows from Eq.~(\ref{FF'}) that the distribution is
given by (\ref{FS}), both in the time direction and in the
spacelike directions along the surfaces $t={\rm const}$. We can
expect, therefore, that the 4-dimensional distribution of $\phi$
among the cells, that is, the distribution in a randomly picked
$4D$ volume, is still given by (\ref{FS}). Note that the form of
the distribution does not depend on the choice of the spacelike
hypersurface $t=0$. This indicates that the distribution is not
sensitive to the details of how the spacetime is divided into
cells, as long as the cell sizes are set by the local horizon.

% Note: this assumes that there is such thing as the "distribution among all 4D cells" but in fact such a thing seems to be not really well-defined and indeed it is gauge-dependent; and we are trying to argue that there is a natural way this gauge dependence can be done away with, and that this natural way is to consider the scale factor time.

\subsection{Ergodic conjecture}

We now turn to the more physically interesting case, when
inflation can end by thermalization in a low-energy vacuum. Here
we shall assume that there is only one such vacuum, so that all
thermalized regions are statistically equivalent. Each thermalized
region has its own infinite hypersurface $\Sigma_q$, which marks
the boundary between quantum diffusion and slow roll. We are
interested in the distribution $P_q(X)$ on $\Sigma_q$. Here, we
have switched back from the notation $\phi_a$ to $(\phi,X)$, where
$X$ stands for one or several light fields.

The simplest guess is that $P_q(X)$ has the same form as the
distribution (\ref{FS}) in the diffusion region, \beq P_q(X)
\propto H^{-2}(\phi_{qX},X)\exp[S(\phi_{qX},X)]. \label{PqS} \eeq
Here, as before, $\phi_{qX}$ is the value of $\phi$ at the onset
of the slow roll for given values of the fields $X$. We do not
have a proof that this guess is correct in general, but there are
some special cases where the distribution (\ref{PqS}) does seem to
apply.

Suppose that the inflaton $\phi$ is trapped in a metastable false
vacuum, $\phi=\phi_F$. The false vacuum decays through bubble
nucleation, which is followed by slow roll and thermalization, as
in models of ``open inflation'' \cite{open}. Prior to bubble
nucleation, the evolution of the fields $X$ is governed by the
stochastic dynamics described in the previous subsection. The
corresponding probability distribution is \beq P(X) \propto
H^{-2}(\phi_{F},X)\exp[S(\phi_{F},X)]. \label{PFS} \eeq

Once a bubble nucleates, the bubble wall expands rapidly with a
speed approaching the speed of light, so
%If we assume that the potential
%energy does not change much across the wall and that the presence of
%the wall does not significantly affect the geometry, then the
the worldsheet of the wall closely follows the future lightcone of the
nucleation point.
%in the unperturbed spacetime.
The onset of slow roll surface $\Sigma_q$ will follow the same
lightcone on the interior side of the bubble.  If we imagine the
spacetime divided into horizon-size cells with different values of
$X$, we can expect that the bubble will cut through a
representative sample of cells and that the distribution of $X$ on
$\Sigma_q$ will be given by (\ref{PFS}). The distribution on the
thermalization surface $\Sigma_*$ is obtained by multiplying
(\ref{PFS}) with the slow roll volume expansion factor, as in
Eq.~(\ref{Pq}).

It should be understood that if the fields $X$ interact with the
bubble, then the distribution (\ref{PFS}) is processed as it goes
through the bubble walls. This effect should be incorporated, just
as the evolution of the fields during slow roll is incorporated in
the Jacobian (\ref{jacobian}). In certain cases, the result of
this processing can be easily estimated. For fields $X$ whose mass
is much smaller than the inverse size of the bubble at the time of
nucleation, the distribution inside the bubble takes the form
\cite{quasiopen} \beq P_q(X) \sim e^{-I_B(X)}. \label{quasiopen}
\eeq Here, the subindex $q$ is used to denote that this is the
distribution at the beginning of slow roll inside the bubble,
right after nucleation. $I_B(X)$ in the exponent is the action of
the Coleman-de Luccia instanton describing bubble nucleation, and
an adiabatic approximation is used for the Euclidean solution,
where the field $X$ is assumed to take a constant value. In fact,
all values of $X$ are realized at distant places inside any given
bubble, but their distribution is still given by Eq.~(\ref{quasiopen})
\cite{quasiopen}. Note that in the limit where there is no bubble,
the Euclidean action is minus the entropy of de Sitter space,
$I_B=-S$, and the expression (\ref{quasiopen}) reduces to
(\ref{PFS}).

In cases where both $\phi$ and $X$ fields are in the diffusion
regime prior to the slow roll, the situation is less clear.
However, the following heuristic argument suggests that the
distribution (\ref{PqS}) may still approximately apply.

Consider an ensemble of comoving observers who start their
evolution in regions where the fields $(\phi,X)$ are near the top
of their potential. Most of these observers will get to the
slow roll and thermalization after a relatively short period of
diffusion. Only a small fraction of the observers will stay in the
diffusion regime for much longer, but their respective regions
will be expanded by a huge factor. This is the essence of eternal
inflation: the histories of these ``atypical'' observers are, in a
certain sense, more representative than those of the typical ones
(after all, the thermalization surface is infinite thanks to the
contribution of regions which linger in the diffusion regime for
an indefinite amount of time.)

The measurements made by observers who spend a long time in the
diffusion regime are described by the FP equation (\ref{FP}),
supplemented by the constraint that the boundary $S_q$ between the
diffusion and slow roll regions in the field space cannot be
crossed. [The surface $S_q$ is defined by the condition
(\ref{phiq}).] Mathematically, this can be enforced by imposing
the reflecting boundary condition on $S_q$. The solution of the FP
equation is then given by the stationary distribution (\ref{FS}).
This indicates that, during diffusion, the observers will see
ergodic evolution in their respective $h$-regions.

We can thus argue, as in the previous subsection, that
Eq.~(\ref{FS}) describes the $4D$ distribution in the quantum
diffusion region of spacetime, which is in the past of the
hypersurface $\Sigma_q$. Then it appears plausible that the same
distribution may extend, at least approximately, to the boundary
$\Sigma_q$ with the slow roll region. This is the ergodic
conjecture~(\ref{PqS}). It assumes that all quantum states
corresponding to the field values on $S_q$ are equally represented
by $h$-regions on $\Sigma_q$.

It should be noted that $\phi_{qX}$ is defined by Eq.~(\ref{phiq})
only in the order-of-magnitude sense. This results in a
significant uncertainty\footnote{This
uncertainty is not present in models with bubble nucleation
discussed earlier in this subsection.}
in the distribution (\ref{PqS}), due to
the exponential dependence on $S(\phi_{qX},X)$.
A possible way of defining
$\Sigma_q$ more precisely is the following. Suppose first there is
a single inflaton $\phi$ and no other light fields. Surfaces of
constant $\phi$ are spacelike in the slow roll regime, but not
necessarily so in the diffusion regime. For $\phi$ in the
diffusion regime, the surfaces have a complicated topology, while
in the slow roll regime the topology is trivial. The transition
between the two might be sharp, having the character of a phase
transition. We could then define $\phi_q$ to be the critical value
of $\phi$. If there are several fields, the phase transition on
the surfaces $H'/H^2=const$ could be used to define $S_q$.

\subsection{An example}

As an illustration, let us consider the potential
\beq
V(\phi,X)={1\over{2}}m^2(X)\phi^2,
\label{V}
\eeq
where $m(X)$ is a very slowly varying function of the fields $X$.
Then
\beq
H(\phi,X)=(\pi/c)m(X)\phi,
\eeq
where $c=(3\pi/8)^{1/2}$, and Eq.~(\ref{phi*}) gives
$\phi_*\sim 1$.

To make the condition (\ref{phiq}) more precise, we rewrite it as
\beq
H'/H^2(\phi_{qX},X) = k
\label{phiq1}
\eeq
with $k\sim 1$, which yields
\beq
\phi_{qX}=\left({c\over{\pi km(X)}}\right)^{1/2}.
\label{phiq2}
\eeq
We shall assume that $m(X)\ll 1$; then $\phi_*\ll\phi_q$.

The ergodic factor in Eq.~(\ref{PqS}) is given by
\beq
\exp[\pi/H^2(\phi_{qX},X)]=\exp[ck/m(X)],
\label{ergm}
\eeq
and the volume expansion factor is
\beq
Z^3\approx\exp(6\pi\phi_{qX}^2)=\exp[6c/km(X)].
\label{Z3}
\eeq
We see that both factors are sensitive to the value of $k$ chosen to
define $\phi_{qX}$. On the other hand, both factors have the form
$\exp[C/m(X)]$ with $C\sim 1$, so we can be reasonably confident that
the full distribution has this form, with a strong peak at the
smallest allowed value of $m(X)$.

\section{Counting pockets}

In models with pockets of several different types, we need to
consider an additional factor characterizing the relative abundances of different types of pockets.  Consider, for instance, a false vacuum which can decay into true vacua 1 or 2, with respective tunneling rates $\kappa_1$ and $\kappa_2$. If $\kappa_1 \gg \kappa_2$, it seems intuitively clear that there will be more pockets of type 1 than of type 2 in the multiverse. This notion, however, should be made more precise.

In this section we propose a method for determining the relative
abundances $p_j$ of different types of pockets. The counting is
performed at the future boundary of spacetime---the place where
everything has been said and done. Of course, the number of
pockets of any given type is infinite, and a cutoff is needed. We
shall argue that one such cutoff suggests itself naturally, and
this allows us to perform an explicit calculation of the $p_j$'s.
One can then examine different examples to check whether this is a
reasonable prescription. As mentioned in the introduction, there
have been in the past alternative proposals for $p_j$ which
nevertheless are not very satisfactory. For comparison, one of
these will be discussed in Section IV.

The full probability distribution should be obtained by combining the weights $p_j$ with the volume distributions $P(X;j)$ in a suitable way. This will be discussed in Section VI.

\subsection{Models with bubble nucleation}

The string theory landscape is expected to have a multitude of
high-energy metastable ``false'' vacua. Such vacua can decay through
bubble nucleation. Bubbles of lower-energy vacuum nucleate and expand
in the high-energy vacuum background. If the ``daughter'' vacuum has a
positive energy density, then inverse transitions are also possible:
bubbles of high-energy vacuum can nucleate in the low-energy one
\cite{EWeinberg}. (This is the so-called ``recycling'' process.) In
both cases, the radius of the bubbles asymptotically approaches the
comoving horizon size in the parent vacuum at the moment of nucleation
\cite{recycling}.  In general, we will have multiple bubbles within
bubbles within bubbles of many different types. The endpoints of this
evolution are the negative or zero-energy vacua, which do not
recycle. We shall call them terminal vacua.

Let $f_j(t)$ be the fraction of comoving volume occupied by vacuum
$j$ at time $t$. The evolution equations for $f_j(t)$ can be
written as \beq {df_j\over{dt}}=\sum_i (-\kappa_{ij}f_j +
\kappa_{ji}f_i), \label{dfdt} \eeq where the first term on the
right-hand side accounts for loss of comoving volume due to
bubbles of type $i$ nucleating within those of type $j$, and the
second term reflects the increase of comoving volume due to
nucleation of type-$j$ bubbles within type-$i$ bubbles. The
transition rate $\kappa_{ij}$ is defined as the probability per
unit time for an observer who is currently in vacuum $j$ to find
herself in vacuum $i$. It can be expressed as \beq
\kappa_{ij}=\lambda_{ij}{4\pi\over{3}}H_j^{\beta-4}, \label{kappa}
\eeq where $\lambda_{ij}$ is the bubble nucleation rate per unit
physical spacetime volume and $H_j$ is the expansion rate in
vacuum $j$. At this point, it will be convenient to distinguish
between the inflating vacua, which we will label by Greek letters,
and the terminal ones, for which we will reserve the indices $m$
and $n$. Then, by definition, \beq \lambda_{\alpha
m}=\lambda_{mn}=0. \label{lambda=0} \eeq

We are interested in the number $N_j$ of nucleated bubbles of a
given low energy vacuum $j$ (not necessarily a terminal one). As
we know, the number of bubbles of all kinds grows without bound,
even within a region of a finite comoving size. We thus need to
cut off our count. The method we propose is to include only
bubbles greater than some small comoving size $\epsilon$. We will
take the limit $\epsilon\to 0$ in our final result. Note that this
cutoff prescription is independent of time parametrization. The
comoving bubble size is set by the horizon at the moment of
nucleation and remains constant in time. Our bubble counting can
therefore be performed at future infinity. We shall call this the
comoving horizon cutoff (CHC) method.

We can relate the number of bubbles nucleated in an infinitesimal
time $dt$ in the parent vacuum $\alpha$, $dN_{j\alpha}(t)$, to the
accompanying increase in comoving volume
$df_{j\alpha}=\kappa_{j\alpha}f_\alpha dt$ (which is the product
of the number of bubbles and the comoving volume each bubble
covers): \beq dN_{j\alpha}(t){4\pi\over{3}}H_\alpha^{-3}a^{-3}(t)
= \kappa_{j\alpha}f_\alpha dt. \eeq Here, $H_\alpha$ is the Hubble
constant of the parent vacuum, $a(t)$ is the scale factor, and we
have used the fact that the comoving radius of bubbles of any type
$j$ nucleating in vacuum $\alpha$ at time $t$ is \beq
R_{j\alpha}(t) = H_\alpha^{-1} a^{-1}(t). \label{R(t)} \eeq It
will be convenient to use the scale factor time, $\beta=0$. Then
$a(t)=e^t$ and \beq
{dN_{j\alpha}(t)\over{dt}}={3\over{4\pi}}H_\alpha^3
e^{3t}\kappa_{j\alpha}f_\alpha. \label{dNdt} \eeq

The comoving volume fractions $f_{j}(t)$ can be found from
Eq.~(\ref{dfdt}), which can be written in a matrix form, \beq
{d{\bf f}\over{dt}}={\mathbf M}{\bf f}, \label{matrixf} \eeq where
${\bf f(t)}\equiv \{ f_j(t)\}$ and \beq
M_{ij}=\kappa_{ij}-\delta_{ij}\sum_r \kappa_{ri}. \label{Mij} \eeq
%The matrix $M_{ij} has the properties
%\beq
%M_{ij}\geq 0 ~~~ (i\neq j),
%\label{prop1}
%\beq
%\sum_i M_{ij}=0.
%\label{prop2}
%\eeq
%The latter property ensures the conservation of probability
%\beq
%{d\over{dt}}\sum_i f_i=0.
%\eeq
The asymptotic solution of (\ref{matrixf}) at large $t$ has the
form \beq {\bf f}(t)={\bf f^{(0)}}+{\bf s}e^{-q t}+ ...
\label{asympt} \eeq Here, ${\bf f}^{(0)}$ is a constant vector
which has nonzero components only in terminal vacua, \beq
f_\alpha^{(0)}=0. \label{falpha=0} \eeq It is clear from
Eq.~(\ref{lambda=0}) that any such vector is an eigenvector of the
matrix ${\mathbf M}$ with zero eigenvalue, \beq {\mathbf M}{\bf
f_0}=0. \eeq As shown in Appendix B, all other eigenvalues of
$\mathbf{M}$ have a negative real part, so the solution approaches
a constant at late times. We have denoted by $-q$ the eigenvalue
with the smallest (by magnitude) negative real part and by
$\mathbf{s}$ the corresponding eigenvector. It will also be shown
in Appendix B that this eigenvalue is real and nondenegerate. This
follows under the assumption that the set of inflating
(nonterminal) vacua is ``irreducible," i.e.~it cannot be split
into groups of noninteracting vacua, where each vacuum from one
group never nucleates any vacua from other groups. If there were
such noninteracting groups, the present considerations would still
apply to each group separately.

The asymptotic values of the terminal components $f_m^{(0)}$
depend on the choice of initial conditions. For any physical
choice, we should have $f_m^{(0)} \geq 0$. Moreover, since $f_m$
can only increase with time, we must have \beq s_m \leq 0.
\label{monk} \eeq At the same time, we should have \beq
s_\alpha\geq 0,\label{monkey} \eeq because $f_\alpha(t)\geq 0$ and
$f_\alpha^{(0)}=0$. (Here, Eqs.~(\ref{monk}) and (\ref{monkey})
are justified on physical grounds. In Appendix B, these equations
will be derived more rigorously.)

It follows from Eqs.~(\ref{asympt}), and (\ref{falpha=0}) that
$f_\alpha \approx s_\alpha e^{-q t}.$ Substituting this into
(\ref{dNdt}) and integrating over $t$, we obtain \beq
N_{j\alpha}(t)= {3\over{4\pi}}H_\alpha^3
{1\over{3-q}}\kappa_{j\alpha} s_\alpha e^{(3-q)t}. \label{N(t)}
\eeq The next step is to impose the cutoff. Our prescription is to
include only bubbles of comoving size greater than $\epsilon$. For
bubbles of type $j$ nucleating in vacuum $\alpha$, this means that
only bubbles nucleated prior to 
\begin{equation} t_{j\alpha}^{(\epsilon)} = -\ln
(\epsilon H_\alpha) \label{t cutoff}\end{equation}
should be included. With this cutoff,
Eq.~(\ref{N(t)}) gives \beq N_j = \sum_\alpha
N_{j\alpha}(t_{j\alpha}^{(\epsilon)}) =
{3\over{4\pi}}{1\over{3-q}}\epsilon^{-(3-q)}\sum_\alpha H_\alpha^q
\kappa_{j\alpha} s_\alpha. \label{Nepsilon} \eeq As expected,
$N_j\to\infty$ as $\epsilon\to 0$. Our proposal is that $p_j
\propto N_j$; hence, \beq p_j\propto \sum_\alpha H_\alpha^q
\kappa_{j\alpha} s_\alpha. \label{pJaume} \eeq The problem of
calculating $p_j$ has thus been reduced to finding the dominant
eigenvalue $q$ and the corresponding eigenvector ${\bf s}$.

We emphasize that our cutoff procedure is independent of
coordinate transformations at future infinity. Any smooth change of
coordinates will locally be seen as a linear transformation, which
amounts to a constant rescaling of bubble sizes, accompanied by a
rotation. Rescaling is generally different in different directions, so
the shapes of the bubbles will be distorted from (approximate) spheres
to ellipsoids. However, in a sufficiently small neighborhood of
comoving size $\delta$, all types of bubbles are distorted in the same
way, so the bubble counting should not be affected by the
distortion. (For a non-spherical bubble, the bubble size can be
defined as its maximum extent, or the major axis of the ellipsoid.)
The bubble density is dominated by the bubbles which formed at very
late times, and which therefore have a very small comoving
size. Hence, in any neighborhood of size $\delta$ the relative numbers
of bubbles will not be affected by constant rescalings in the limit
$\epsilon\to 0$.  This means that calculation of $p_j$ in any such
neighborhood will give the same result, and it is clear that this
result will also hold in a comoving region of any size.

\subsection{Models with quantum diffusion}

In quantum diffusion models, the calculation can be performed in a
similar way, except the role of bubbles is now played by newly
thermalized $h$-regions. The rate at which the comoving volume
thermalizes can be calculated by using the FP equation.

The FP equation for the distribution function $F(\phi,t)$ applies in
the region of $\phi$-space outside the
thermalization boundaries $S_*^{(j)}$. The boundary condition on $F$
requires that diffusion vanish on $S_*^{(j)}$ \cite{LLM94}. In the notation of
Sec. II.B,
\beq
{\hat n}_a{\p\over{\p\phi_a}}[D(\phi)F]=0 ~~~~ (\phi\in S_*^{(j)}),
\label{bc}
\eeq
where ${\hat n}_a$ is the normal to $S_*^{(j)}$.

The fraction of comoving volume which ends up in a thermalized
region of type $j$ per unit time with the fields $\phi_{a*}$ in an
infinitesimal surface element $dS_*^{(j)}$ of the thermalization
boundary $S_*^{(j)}$ is given by \beq
 {\hat n}_a(\phi_*) J_a(\phi_*,t) dS_*^{(j)}.
\label{PJ1}
\eeq
Using
Eqs.~(\ref{J}) and (\ref{v}) and the boundary condition (\ref{bc}), this
can be expressed as
\beq
{1\over{4\pi}} \left|{\p H\over{\p n}}(\phi_*)
\right| H^{\beta-1}(\phi_*)F(\phi_*,t) dS_*^{(j)} ,
\label{Ppsi1}
\eeq
where $(\p H/\p n)\equiv {\hat n}_a (\p H/\p \phi_a)$.

The fraction of comoving volume that thermalizes during the scale
factor time interval $dt$ ($\beta=0$) is thus given by \beq
{1\over{4\pi}}H^{-1}(\phi_*)\left|{\partial H\over{\partial
n}}(\phi_*)\right| F(\phi_*,t) dS_* dt. \eeq The asymptotic
comoving size of a pocket universe $r_j$ is much bigger than the
comoving size of an $h$-region at the time of thermalization. Let
us denote by $\lambda_*$ the ratio of these two sizes, \beq
r_j\equiv \lambda_* H_*^{-1} e^{-t_*} \label{cosi} \eeq where
$t_*$ is the time of thermalization. In fact, $r_j$ is of the
order of the comoving size of the horizon at the time $t_q$ when
slow roll begins, which is of order $H_q e^{-t_q}$.
Thus,\footnote{Regions in the vicinity of an $h$ region which has
just entered slow roll, have field values which are also close to
the boundary where slow roll begins, and so they are likely to
cross it soon after. Because of such correlations, we expect that
$r_j$ is somewhat larger than the comoving horizon size at the
time $t_q$, by a factor of a few, which can be determined from
numerical simulations.} \beq \lambda_* \sim (H_*/H_q)e^{(t_*-t_q)}
\sim (H_*/H_q) Z, \label{lambdap} \eeq where $Z$ is approximately
given by (\ref{Z}). The number of pocket universes which begin
thermalizing at time $t$ then satisfies \beq
dN_j(\phi_*,t)={3H^2(\phi_*)\over{(4\pi)^2}}
\frac{e^{3t}}{\lambda_*^{3}} \left| {\partial H\over{\partial
n}}(\phi_*)\right| F(\phi_*,t) dS_* dt. \label{dN} \eeq

The asymptotic solution of the FP equation has the form \beq
F(\phi,t)\approx F_1(\phi)\exp(-q_1 t), \eeq where $-q_1$ is the
smallest (by magnitude) eigenvalue of the FP operator.
Substituting this in (\ref{dN}) and integrating over $t$, we find
the number of $h$-regions that have thermalized up to time $t$
with $\phi_*$ in a given surface element $dS_*$. The probabilities
$p_j$ are then found after a cutoff at
$t^{(\epsilon)}(\phi_*)=-\ln [\epsilon H(\phi_*)/\lambda_*]$ and
integration over $\phi_*$, \beq p_j \propto \int_{S_*^{(j)}}dS_*
\lambda_*^{-q_1} H^{-(1-q_1)}(\phi_*) \left| {\partial
H\over{\partial n}}(\phi_*)\right| F_1(\phi_*). \label{pJaume2}
\eeq This equation allows us to calculate $p_j$ once we find the
dominant eigenvalue $q_1$ and the corresponding eigenfunction
$F_1(\phi)$. It should be noted that in the slow roll region, the
diffusion term can be neglected and the Fokker-Planck equation
becomes first order. One then readily finds that $J_\phi \propto
Z^q(\phi)$, where $Z$ is the slow roll expansion factor. With
$\lambda_*$ given by (\ref{lambdap}), it is clear that the $p_j$
do not depend on where exactly we choose to define the
thermalization boundary. The integrand can be evaluated anywhere
in the slow roll regime, as long as diffusion is negligible. In
fact, Eq.~(\ref{pJaume2}) is approximately valid if we substitute
the subindex $*$, corresponding to thermalization, by the subindex
$q$, indicating the onset of the slow roll regime. In this case
$\lambda_q \sim 1$. Note that the value of $q_1$ is typically very
small.

In the preceding subsection we assumed that the physics of eternal
inflation is described by bubble nucleation, and in the present
subsection we assumed that it is described by quantum diffusion. The
general case, when both mechanisms are present, can be described by
combining the formalisms outlined in these subsections
\cite{recycling}.

\section{Comoving probability}

An alternative weight factor for the different types of pockets,
$p^c_j$, was used in Ref.~\cite{GV01}. This is the so-called
comoving probability, which can be defined as the probability for
a comoving observer, starting near the top of the potential
$V(\phi,X)$, to end up in a pocket of type $j$. Let us now briefly
describe, for comparison, the calculation of such probabilities.

\subsection{Models with quantum diffusion}

%The boundary $S_*$ can be divided into segments $S_*^{(j)}$
%corresponding to the attraction basins of different minima of the
%potential, which we label by the index $j$. Our goal is to calculate
%the probabilities $P_j$ for these minima.

The form of the FP equation depends on the time parametrization
parameter $\beta$, and the resulting distribution $F(\phi,t)$ is, of
course, also $\beta$-dependent. However, as shown in \cite{AV99}, this distribution can be used to define
reparametrization-invariant probabilities.

The probability for a comoving observer to end up in a thermalized region
of type $j$ is given by
\beq
p^c_j=\int_{S_*^{(j)}}dS_* {\hat n}_a(\phi_*)\int_0^\infty dt
J_a(\phi_*,t),
\label{PJ}
\eeq
where the first integration is over $\phi_*\in S^{(j)}_*$. Again, using
Eqs.~(\ref{J}) and (\ref{v}) and the boundary condition (\ref{bc}), this
can be expressed as
\beq
p^c_j={1\over{4\pi}}\int_{S_*^{(j)}}dS_* \left|{\p H\over{\p n}}(\phi_*)
\right|\psi(\phi_*),
\label{Ppsi}
\eeq
where we have introduced
\beq
\psi(\phi)=H^{\beta-1}(\phi)\int_0^\infty dt F(\phi,t).
\label{psi}
\eeq
Integrating the FP equation (\ref{FP}),(\ref{J}) and the boundary
condition (\ref{bc}) over time, we obtain the corresponding equations
for $\psi(\phi)$:
\beq
{\p\over{\p\phi_a}}\left[{1\over{8\pi^2}}{\p\over{\p\phi_a}} (H^3\psi)
+{1\over{4\pi}}{\p H\over{\p\phi_a}}\psi \right]=-F_0(\phi),
\label{FPpsi}
\eeq
\beq
{\p\over{\p n}}(H^3\psi)=0 ~~~~ (\phi\in S_*^{(j)}).
\label{bcpsi}
\eeq
Here, $F_0(\phi)=F(\phi,0)$ is the initial distribution at $t=0$.

The function $\psi(\phi)$ is uniquely determined by Eq.~(\ref{FPpsi})
with the boundary conditions (\ref{bcpsi}). It can then be used in
Eq.~(\ref{Ppsi}) to evaluate the probabilities $p_j^c$. Note that the
parameter $\beta$ has been absorbed in the definition (\ref{psi}) of
$\psi(\phi)$ and does not appear in
Eqs.~(\ref{Ppsi}),(\ref{FPpsi}),(\ref{bcpsi}). This shows that this
definition of $p_j^c$ is independent of time parametrization.

The physical origin of this reparametrization invariance is easy
to understand. If we start with a large inflating volume at $t=0$,
different parts of this volume will thermalize into different
minima of the potential, and at any time $t$ there will be parts
of this volume that are still inflating. However, in the limit
$t\to\infty$ all the comoving volume will be thermalized, except a
part of measure zero. Different choices of time parametrization
affect the division of the volume into thermalized and still
inflating regions, but this has no effect on $p_j^c$, since the
inflating regions represent only an infinitesimal part of the
comoving volume at $t\to\infty$.  In other words, the fraction of
comoving volume that ends up in a given minimum of the potential
is gauge-independent.

\subsection{Models with bubble nucleation}

Likewise, in models with bubble nucleation it is convenient to
introduce the new variable \beq \psi_j =
{4\pi\over{3}}H_j^{\beta-4}\int_0^\infty f_j(t)dt \label{psij}.
\eeq Integrating Eq.~(\ref{dfdt}) over time, we obtain \beq
f_j(\infty)-f_j(0)=\sum_k (-\lambda_{kj}\psi_j +
\lambda_{jk}\psi_k) \label{fpsi} \eeq In the asymptotic future,
all the comoving volume will be in the terminal vacua. Hence, \beq
f_\alpha(\infty)=0 \label{finfty} \eeq for all non-terminal
vacua, whereas the probabilities for terminal vacua are given by
\beq p_n^c = f_n(\infty). \label{p=f} \eeq On the other hand, the
initial distribution is concentrated at high-energy vacua; hence
\beq f_n(0)=0. \label{f0} \eeq

Now, it follows from the above equations that \beq \sum_\alpha
(\lambda_{\alpha\gamma}\psi_\gamma - \lambda_{\gamma\alpha}
\psi_\alpha) = f_\gamma(0) \label{alpha} \eeq and \beq p_n^c =
\sum_\alpha \lambda_{n\alpha}\psi_\alpha. \label{j} \eeq The
quantities $\psi_\alpha$ can be determined from (\ref{alpha}), and
the probabilities $p_n^c$ can then be found from Eq.~(\ref{j}). As
before, the time reparametrization parameter $\beta$ has been
absorbed in the definition of $\psi_\alpha$, and the probabilities
do not depend on the choice of the time variable.

\subsection{Problems with $p^c_j$}

The probabilities $p_j^c$ in (\ref{Ppsi}) and (\ref{j}) are independent
of time parametrization, but they do depend on the initial probability
distribution $F_0(\phi)$ or $f_\alpha(0)$. We assumed that the initial
distribution is concentrated near the maximum of the potential, or in
the highest-energy false vacuum. However, if the potential has several
peaks of comparable height, different values of $p_j^c$ will be obtained
starting from different peaks.

It is possible that the initial distribution is to be found from the wave
function of the universe, which determines the probability distribution for
the initial states of the universe as it nucleates out of nothing. It is
well known that, although inflating spacetimes are generically eternal to
the future, all past-directed geodesics in such spacetimes are incomplete,
except perhaps a set of measure zero (see \cite{BGV} and references
therein). This indicates that the inflating region of spacetime has a
boundary in the past, and some new physics (other than inflation) is
necessary to determine the initial conditions at that boundary. The prime
candidate for the theory of cosmic initial conditions is quantum cosmology,
which suggests that the universe starts as a small, closed 3-geometry and
immediately enters the regime of eternal inflation.

If, for example, we adopt the tunneling wave function of the
universe, the initial distribution is given by \cite{AV84,Linde84}
\beq
F_0(\phi)\propto \exp\left(-{3\over{8V(\phi)}}\right). \label{qc}
\eeq This distribution favors large values of $V(\phi)$. So, if
the potential is dominated by a single peak, the initial
distribution will be concentrated at that peak.

The distribution (\ref{qc}) may be a plausible choice for the
initial state of the universe, but the conclusion that the
probabilities of different observations in an eternally inflating
universe have some dependence on this choice appears to be
counter-intuitive. Even though inflation must have had a
beginning, one expects that once it began, the universe will
quickly forget its initial conditions.

Another problem is that, from (\ref{finfty}), \beq p^c_\alpha =0,
\label{drawback} \eeq for all non-terminal vacua. If we use
$p^c_j$ for determining the probabilities of being in different
pockets, all non-terminal vacua are given zero probability.
However, there seems to be no good reason to discard all such vacua. Note that the relative abundance of the corresponding pockets, discussed in the previous section, can be sizable. Also, our own low energy vacuum may have a small positive cosmological constant, in which case it would be non-terminal. This seems to disfavor the use of $p^c_j$ as a relevant weight factor.

The weight factors $p_j$ which we discussed in Section III do not suffer from these problems, and therefore seem more suitable for our present purposes.

\section{Some examples}

We shall now use some examples to illustrate the similarities and
differences between the weight factors $p_j$ and $p_j^c$. Consider first a
very simple model with one false vacuum $F$ and two terminal
vacua $A$ and $B$ ($FAB$ model). The allowed transitions in this model
are shown by the ``schematic'' \beq A\leftarrow F\rightarrow B.
\label{AFB}
\eeq
The vector ${\bf f}$ then has three components, ${\bf
f}\equiv (f_F,f_A,f_B)$, and the evolution equations (\ref{dfdt})
have the form
\beqa
\frac{df_F}{dt}=
-(\kappa_{AF}+\kappa_{BF})f_F, \label{evolFAB0} \\
\frac{df_A}{dt}= \kappa_{AF}f_F, \label{evolFAB1} \\
\frac{df_B}{dt}= \kappa_{BF}f_F.
\label{evolFAB}
\eeqa

The first of these equations gives
\beq
f_F(t)=Ce^{-qt},
\eeq
with $C={\rm const}$ and
\beq
q=\kappa_{AF}+\kappa_{BF}.
\label{qforFAB}
\eeq
With the initial conditions ${\bf f}(0)=\{1,0,0\}$, we have $C=1$, and
the other two equations yield
\beq
f_j(t)=(\kappa_{jF}/q)\left( 1-e^{-qt}\right),
\eeq
where $j=A,B$. The comoving probabilities are defined by $p_j^c\propto
f_j(\infty)$, which gives
\beq
p_j^c\propto \kappa_{jF}.
\label{pjFAB}
\eeq
This is in accord with the intuitive expectation: the probability of a
vacuum is proportional to the nucleation rate of the corresponding
bubbles.

We now compare this with the CHC (comoving horizon cutoff)
approach. The transition matrix ${\mathbf M}$ corresponding to
Eqs.~(\ref{evolFAB0})-(\ref{evolFAB}) has two zero eigenvalues
with eigenvectors $(0,1,0)$ and $(0,0,1)$, and a negative
eigenvalue $-q$ [where $q$ is given by Eq.~(\ref{qforFAB})] and
eigenvector ${\bf s}=(1,-\kappa_{AF}/q,-\kappa_{BF}/q)$.  For our
$FAB$ model, the sum in Eq.~(\ref{pJaume}) has only one term (the
one with $\alpha=F$), and the only $j$-dependent factor in
(\ref{pJaume}) is $\kappa_{jF}$. Thus, the CHC prescription gives
\begin{equation} p_j\propto \kappa_{jF},
\end{equation}
in agreement with Eq.~(\ref{pjFAB}).

We next consider a model with an intermediate vacuum $I$ between $F$
and $B$,
\beq
A\leftarrow F\rightarrow I\rightarrow B
\eeq
($FABI$ model). Bubbles of $A$ and $I$ nucleate in $F$, but bubbles of
$I$ are themselves sites of eternal inflation, with an infinite number
of $B$-bubbles nucleating in each of them. This model illustrates the
difference between the weight factors $p_j$ and $p_j^c$.

The evolution equations for the $FABI$ model are
\beqa
\label{evoFABI}
\frac{df_F}{dt}=
-(\kappa_{AF}+\kappa_{IF})f_F, \\
\frac{df_I}{dt}= -\kappa_{BI}f_I+\kappa_{IF}f_F, \\
\frac{df_A}{dt}= \kappa_{AF}f_F, \\
\frac{df_B}{dt}= \kappa_{BI}f_I.
\label{evolFABI}
\eeqa
Once again, we start with the comoving probabilities. The solution of
Eqs.~(\ref{evoFABI})-(\ref{evolFABI}) with the initial conditions
$(f_F,f_I,f_A,f_B)(t=0) = (1,0,0,0)$ is
\beqa
f_F(t)=e^{-qt}, \\
f_I(t)=C\left( e^{-qt}-e^{-\kappa_{BI}t}\right), \\
f_A(t)={\kappa_{AF}\over{q}}\left( 1-e^{-qt}\right), \\
f_B(t)={\kappa_{IF}\over{q}}+C\left( e^{-\kappa_{BI}t}
-{\kappa_{BI}\over{q}} e^{-qt}\right),
\label{f(t)FABI}
\eeqa
where
\beq
q=\kappa_{AF}+\kappa_{IF}
\label{qFABI}
\eeq
and
\beq
C={\kappa_{IF}\over{\kappa_{BI}-q}}.
\eeq
The comoving probabilities for this model are
\beq
{p_A^c\over{p_B^c}}={f_A(\infty)\over{f_B(\infty)}} =
{\kappa_{AF}\over{\kappa_{IF}}}.
\label{pApBFABI}
\eeq

This result is easy to understand. We start with all comoving volume
in $F$. This volume is then divided between the bubbles of $A$ and $I$
in the ratio $\kappa_{AF}/\kappa_{IF}$. All the comoving volume in
$I$-bubbles is eventually turned into $B$-bubbles, and thus the ratio
of comoving volumes in $A$ and $B$ is given by (\ref{pApBFABI}).

Let us now compare this with the CHC method. The matrix
${\mathbf M}$ for the $FABI$ model has two nonzero eigenvalues:
$-q$ with $q$ from (\ref{qFABI}) and $-q'$ with $q'=\kappa_{BI}$. The
corresponding eigenvectors are
\beq
{\bf s}=(s_F,s_I,s_A,s_B)=\left({q(q-q')\over{\kappa_{IF}q'}}, -{q\over{q'}},
-{\kappa_{AF}(q-q')\over{\kappa_{IF}q'}}, 1\right)
\label{sFABI}
\eeq
and
\beq
{\bf s'}=(0,-1,0,1).
\label{s'FABI}
\eeq
The asymptotic behavior of the model depends on the relative magnitude
of $q$ and $q'$.

For $q<q'$, or
\beq
\kappa_{AF}+\kappa_{IF}<\kappa_{BI},
\label{q<q'}
\eeq
$q$ is the dominant eigenvalue, and Eq.~(\ref{pJaume}) gives
\beq
{p_A\over{p_B}}=\left({H_F\over{H_I}}\right)^q
{\kappa_{AF}(q'-q)\over{\kappa_{IF}q'}}.
\label{JaumeFABI}
\eeq
Conversely, if $q>q'$, then $q'$ is the dominant eigenvalue, and we
find
\beq
{p_A\over{p_B}}=0.
\eeq

Once again, these results are easy to understand. For $q>q'$, the
high-energy vacuum $F$ decays faster than the intermediate vacuum
$I$. So, in the asymptotic regime the comoving volume is divided
between bubbles of $A$ and $I$, and each bubble of $I$ still continues
to produce bubbles of $B$. As a result, $A$-bubbles are completely
outnumbered by $B$-bubbles. In the opposite limit, when $q\ll
q'\lesssim 1$, $I$ turns into $B$ very quickly, and in this sense the
model is not very different from the $FAB$ model. In this limit,
Eq.~(\ref{JaumeFABI}) gives $p_A/p_B \approx
\kappa_{AF}/\kappa_{IF}$, in agreement with (\ref{pApBFABI}).

It is not difficult to extend this analysis to the case where the
intermediate vacuum $I$ is recyclable, so the schematic is
$A\leftarrow F\leftrightarrow I\rightarrow B$. This is done in
Appendix A.

We thus see that both weight factors $p_i$ and $p_i^c$ give
results which are in agreement with intuitive expectations. A
priori, it is not clear which one of these two objects is more
useful for the purposes of defining probabilities for the
constants of nature. Progress can be made by working out their
values in a variety of models, to see which, if any, of the two
can be included in a reasonable definition of probabilities. As
discussed in the previous section, it appears that on general
grounds the $p_i$ are preferable. This is because of their
independence of the initial conditions, and because they assign
non-vanishing probabilities to non-terminal vacua (unlike the
co-moving probabilities). Also, the example considered in this
section indicates that in models where there are intermediate
vacua which are also eternally inflating, the probabilities $p_i$
seem to better represent the actual distribution of terminal
pockets $A$ and $B$. Hence, in what follows, we shall concentrate
on $p_i$.

\section{The full distribution}

Let us now address the question of calculating probabilities for
observations in the case when there are different types of
pockets. The full probability distribution may be written as \beq
{P}_{obs}(X)\propto \sum_{j}
P_j(X;t_j)n_{obs}^{(j)}(X;t_j). \label{calP} \eeq where
$n_{obs}^{(j)}(X;t_j)$ is defined as the number of observers that will
evolve per unit comoving volume with specified values of the
fields $X$, and $t_j$ is a local time variable in a pocket of
type $j$. The calculation of $n_{obs}^{(j)}(X;t_j)$ will not be
discussed in this paper, and we shall simply assume that, given the
reference class of observers, this can be calculated from first
principles.  The distribution $P_j(X;t_j)$ should represent the
physical volume fraction in pockets of type $j$, with specified values
of the fields $X$. Within a given pocket, the quantities $P_j(X;t_j)$
and $n_{obs}(X;t_j)$ should be calculated at the same time $t_j$,
but it is
not important which moment of time and which time variable we choose. The
volume grows and the density decreases with the expansion, but of
course the product does not change. (Note that
$n_{obs}^{(j)}(X;t_j)$ includes all observers---present, past, and
future---that will ever evolve in a comoving volume having unit size
at $t_j$, so its $t$-dependence is simply $a_j^{-3}(X;t_j)$, where
$a_j(X;t)$ is the scale factor.)

In Ref.~\cite{GV01} an attempt was made to compare the volumes in
different pockets. The idea was to quantify the amount of expansion
measured from the inflaton's last visit to the highest point in its
potential. For that purpose, some fictitious markers were introduced,
which were produced at a constant rate in regions where the field is
at the top of the potential, and which where subsequently diluted. The
mean separation of the markers in the different thermalized regions
would then indicate how much expansion intervened since the field was
at the top. This method runs into some conceptual problems. If there
are several peaks of comparable height in the potential, then it is
not clear why we should only count one of them as the source of
markers. Also, since markers are only produced at a very special field
value, all field values which are distant from it get rewarded with a
large expansion factor, for no particularly compelling reason.  In the
face of such difficulties, one option would be to content ourselves
with the volume distributions for individual pockets as the last
frontier of predictivity.

Nevertheless, the structure of the eternally inflating
spacetime is well understood, and we feel that a useful
characterization of the relative likelihood the constants of
nature in pockets of different types can be obtained from the
distributions $p_j$ and $P(X;j)$ which we have considered in the
previous sections.

Let us start, for simplicity, with the case where the different
pockets are generated by bubble nucleation. Suppose also that
different pockets have different values of the constants, but that
these constants do not vary within a given bubble. Our goal is to
determine the probability of being in a given pocket $j$,
$P_{obs}^{(j)}$. In this case, the problem is that of combining
$p_j$ and $n_{obs}^{(j)}$ into $P_{obs}^{(j)}$ [for the time
being,  there is no $P(X;j)$ to worry about, since the constants
$X$ do not vary within a given pocket].

For that, we need to define a co-moving reference scale specifying
the size of the regions in which the observers are to be counted,
and we can write
\beq
P_{obs}^{(j)}  \propto  p_j R_j^3(t) n_{obs}^{(j)}(t),
\eeq
where $R_j(t)$ is the physical size of the reference scale and we
have omitted the subscript $j$ of $t_j$. To determine $R_j(t)$, we
note that at early times, the dynamics of the open FRW universe inside
a bubble is dominated by curvature, and we have $a_j\approx t$ for all
types of bubbles. Later on, when the curvature scale grows to the size
of the Hubble radius associated to the corresponding local energy
density, the curvature dominated phase gives way to slow roll
inflation inside the bubble. For a quasi-de Sitter slow roll phase we
would have
\beq
a_j(t)  \approx (1/H_{q}^{(j)}) \sinh (H_{q}^{(j)} t),
\eeq
where $H_{q}$ is the Hubble rate at the beginning of the slow
roll. The specific form of the scale factor at late times is not
important for our argument. The point is that for times much
smaller than the onset of slow roll, $t \ll 1/H_{q}^{(j)}$, all
pockets are nearly identical, with scale factor $a_j(t) \approx
t$. This suggests that the reference scales should be chosen so
that $R_j(t)$ is the same in all pockets at some $t = \epsilon \ll
1/H_{q}^{(j)}$. Then, up to a constant,
\beq
R_j(t) = a_j(t).
\eeq
For times $t\gg 1/H_q^{(j)}$ this can be expressed as  $R_j(t)
\approx (1/H_{q}^{(j)}) Z_j(t)$, where $Z_j$ is the redshift
factor from the time at which slow roll begins. Alternatively, we
note that $R_j(t)$ is the comoving curvature scale, which can be
defined without any reference to $\Sigma_q$.

Now, this can be straightforwardly generalized to the case when
there are some continuous variables $X$, by simply including the
factor ${\hat P}_\epsilon (X;j)$---the normalized distribution for
$X$ on the surface $t = \epsilon$:
\beq
P_{obs}(X)  \propto  \sum_j p_j {\hat P}_\epsilon(X;j) R_j^3(X)
n_{obs}^{(j)}(X).
\eeq
The volume distribution of $X$ on the surface $\Sigma_q$ is
\beq
{\hat P}_q(X;j)  =  v_j^{-1} {\hat P}_\epsilon(X;j) H^{-3}_j(X), \label{vjdef}
\eeq
where we abreviate $H_j\equiv H_q^{(j)}$, and $v_j$ is the
normalization factor
\beq
v_j =  \int {\hat P}_\epsilon(X;j) H^{-3}_j(X) dS_q  =
\langle H_j^{-3}\rangle _\epsilon.
\eeq
Combining the above equations, the fraction of volume in pockets
of type $j$ with values of the constants $X$ takes the form
\begin{equation}
P_j(X) \propto p_j  v_j \hat P_q(X;j)Z_j^3(X). \label{option3}
\end{equation}
Here, $\hat P_q$ is the single pocket volume distribution
normalized at the onset of slow roll (in the case of diffusion,
this can be estimated as the ergodic factor $P_q(X;j) \sim N
\exp[S(X;j)]$, or using Eq.~(\ref{quasiopen}) in the case of
bubble nucleation). The factor $v_j$ has the dimension of volume. From (\ref{vjdef}) we have $v_j H^3_j \hat P_q = \hat P_\epsilon$. The left hand side of this equation is normalized to unity, and therefore we may write $v_j$ in terms of the volume distribution on $\Sigma_q$,
\beq v_j=
\left[\int H^3_j(X) \hat P_q(X;j) dS_q\right]^{-1}. \eeq Note that
the product $P_h(X;j)\propto H^3_j(X) P_q(X;j)$ featuring in the
denominator of $v_j$ is the number distribution of $h$-regions on
$\Sigma_q$ with the given value of $X$ in pockets $j$. An
alternative form of (\ref{option3}) is given in terms of this
number distribution by
\begin{equation}
P_j(X) \propto p_j  \hat P_h(X;j)H^{-3}_j(X)Z_j^3(X), \label{hreg}
\end{equation}
where $\hat P_h$ is normalized on $S_q$. This result has a simple
intuitive interpretation. The volume distribution is proportional
to the relative number of pockets of a given type $p_j$, times the
relative number of Hubble regions within these pockets with given
values of $X$, $\hat P_h(X;j)$, on the surface where slow roll
starts, times the volume of these regions $H_j^{-3}$, times the
subsequent growth factor $Z_j^3$.

Our derivation of (\ref{option3}), or its equivalent form
(\ref{hreg}), has been based on the case of bubbles, but it is
possible to adopt the same prescription also for the case of
pocket universes generated by diffusion, since the result only
makes reference to the surface $\Sigma_q$ where slow roll begins
and to the subsequent expansion thereof.

At present, we cannot claim that the prescription we have
introduced here is the only consistent one. However, the
hope is that from the careful analysis of a sufficiently wide
range of possibilities a unique prescription will emerge. To
illustrate this point, let us consider two alternatives which on
the surface may seem reasonable.

Let us first consider a possibility which we shall refer to as
Alternative $A$. This consists of identifying $p_j$ with the
volume fraction in pockets of type $j$. In this case, we would
have \beq P_j(X)=p_j {\hat P}(X;j). \label{PjPhat} \eeq Here,
${\hat P}(X;j)$ is the distribution $P(X;j)$ normalized on the
corresponding thermalization boundary, \beq \int dS_*^{(j)} {\hat
P}(X;j)=1. \eeq Although the choice (\ref{PjPhat}) may appear
reasonable at first sight, it is unsuitable for substitution into
Eq.~(\ref{calP}). Indeed, the product $P_j(X)n_{obs}^{j}(X)$
should be independent of the time $t_j$ at which we choose to calculate
it. But with the present choice, $P_j(X)$ stays constant, while
$n_{obs}^{j}(X)$ dilutes with the expansion.

To avoid this problem, we may then think of another possibility,
which we shall refer to as Alternative $B$, where \beq P_{obs}(X)
= \sum p_j \hat P_{obs}^{j}(X).\label{PjPhatb}\eeq  Here, it is
$\hat P_{obs}^{j}(X) = N P(X;j) n_{obs}^{j}$ which is normalized
to one. With this, the time at which we calculate $P(X;j)
n_{obs}^{j}$ is unimportant, since this product is independent of
time. This proposal, however, does not take the form (\ref{calP})
and contradicts our starting point that all members of the
reference class of observers carry equal weight (which is in fact
the very definition of reference class). Suppose we have a single
false vacuum decaying into two types of bubbles with equal
nucleation rates, so that $p_1 = p_2$. The vacuum in type-1
bubbles is very hospitable to observers, while the vacuum in
type-2 bubbles is almost lethal and density of observers is
strongly suppressed. One expects that observers are much more
likely to find themselves in a type-1 bubble. However, after the
normalization prescribed in (\ref{PjPhatb}), we would conclude
that we are equally likely to be in either type of bubble. The
basic problem here is that the factors $p_j$ are determined by the
transition rates, or by the diffusion dynamics, and therefore they
will be insensitive to the amount of slow roll inside of a given
pocket---and much less to the number of observers which will
subsequently develop. Hence, we must also reject this alternative.

The preceeding discussion illustrates that, as we have seen in the
case of Alternatives $A$ and $B$, certain reasonably-looking
prescriptions can be discarded after careful consideration of
their consistency. Another way of narrowing down the possibilities
is by working out their predictions and comparing them with the
data.\footnote{This type of analysis has already proved useful
when applied to definitions of probability based on a
constant-time cutoff. For a wide class of time variables, such
definitions lead to absurd predictions that we should find
ourselves in a deep, nearly spherical gravitational potential well
\cite{LLM2} and that the CMB temperature should be many orders of
magnitude higher than observed \cite{Tegmark}. (As we mentioned in
the Introduction, another objection against constant-time cutoffs
is that they rely on an arbitrary choice of the time variable.)}
Although a full analysis is beyond the scope of the present paper,
we believe that the definition (\ref{option3}) [or equivalently
(\ref{hreg})] is the most appealing amongst those which we have
considered, and has the best chance of being correct.

\section{Summary and discussion}

In the context of theories with many vacua, such as the landscape of string theory, the low energy constants of Nature are randomized during inflation, and will vary from place to place in the thermalized post-inflationary regions.
An interesting question is then to find the volume distribution of the constants in such thermalized regions. In this paper, we have developed some tools towards this goal.

The problem splits into two parts. First of all, we should find
the distribution of constants within a given pocket universe.
Following up on the approach of Refs. \cite{AV98,VVW}, we have
motivated an analytic estimate for this distribution. According to
Eq.~(\ref{Pq}), the internal volume distribution in a given pocket
of type $j$ takes the form \beq P(X;j) \propto P_q(X;j) Z^3(X;j),
\label{PXj} \eeq where $P_q(X;j)$ and $Z(X;j)$ are the
distribution at the onset of slow roll and the slow roll expansion
factor respectively. We argued in Section II.C that \beq
P_q(X;j)\approx H^{-2}(X;j)\exp [S(X;j)]. \label{PqXj} \eeq This
is what we called the ergodic conjecture.\footnote{In models of
bubble nucleation followed by slow roll inside the bubble,
Eq.~(\ref{PqXj}) should be replaced with an expression of the form
(\ref{quasiopen}).} It should be understood that our justification
of (\ref{PqXj}) has been rather heuristic, and the limits of its
validity should be further explored.

Next, we have introduced a weight factor $p_j$ which counts the
relative number of pockets of type $j$. The counting is done at
the future boundary of spacetime, by considering all pockets of
comoving size larger than some $\epsilon$, and then letting
$\epsilon\to 0$. We have called this the comoving horizon cutoff
(CHC), since the comoving size of a pocket is roughly given by the
size of the horizon at the time when the pocket is formed. As we
explained at the end of Section III.A, such a cutoff is independent
of coordinate transformations at the future boundary.

For comparison, we have also discussed an alternative weight
factor $p_j^c$, defined in \cite{GV01} as the fraction of comoving
volume which ends up in pockets of type $j$. Unlike the $p_j$'s
discussed above, the comoving volume fraction $p_j^c$ strongly
depends on initial conditions. Moreover, all co-moving volume ends
up in terminal vacua (these are vacua which cannot further decay
into other vacua). Because of that, all non-terminal vacua are
assigned zero probability. However, there seems to be no good
reason to discard all such vacua, since the relative abundance of
the corresponding pockets can be sizable. Also, our own low energy
vacuum may have a small positive cosmological constant, in which
case it would be non-terminal. Because of these two problems, the
probabilities $p^c_j$ seem less relevant for our present purposes.

In the case where all terminal vacua emanate directly from a
single eternally inflating false vacuum, we find that the weight
factors $p_j$ and $p_j^c$ agree with each other, and they are
basically proportional to the corresponding nucleation rates. The
differences between $p_j$ and $p_j^c$ are illustrated in a model
where there is an intermediate false vacuum which is also
eternally inflating. Again, the weight factors $p_j$ seem to give
a better representation of the actual distribution of pockets in
the multiverse.

We have combined the above distributions into an object
describing the thermalized volume distribution of the constants of nature in
the inflationary multiverse with pockets of different types. If
these constants are determined by the values of certain light
fields $X$, then we have argued that the volume fraction occupied
by values $X$ in pockets of type $j$ is given by 
\beq dP_j(X) =
P_j(X)\ dS^{(j)}_q = p_j\ \hat P_h(X;j)\frac{Z^3_j(X)}{ H_j^3(X)}\ d S^{(j)}_q.
\label{smry}\eeq 
Here, $S_q^{(j)}$ is the boundary (in field
space) where quantum diffusion turns into classical slow roll. The
factor $p_j$, whose calculation is discussed in Section III, takes
into account the relative numbers of pockets of type $j$.  The
quantity $\hat P_h(X;j)$ is the number distribution of 
Hubble-size regions with values of the fields $X$ within pockets of
type $j$ at $\Sigma_q$ (see Fig.~2 for the space-time structure of a
pocket universe),  and $H_j^{-3}(X)$ is the volume of one such
region.  According to the ergodic conjecture, this distribution
is essentially given by Eq.~(\ref{PqXj}) [or Eq.~(\ref{quasiopen}) in the
case of bubbles].  The hat on $\hat P_h$ indicates that 
it is normalized as $\int \hat P_h(X;j) d S_q^{(j)}
=1$. $Z^3_j(X)$ is the expansion factor from the point $X$ on $S_q$ to
the thermalization boundary $S_*$, along the classical slow roll
trajectory.

Although our proposal for comparing the probabilities in
pockets of different types seems to be well motivated, it may not
be the unique possibility. Settling this issue may require further
research. For comparison, we have also explored some alternatives.
In the first one, the weight factor $p_j$ is formally identified
with the volume fraction in pockets of type $j$, and in the second
one, $p_j$ is formally identified with the fraction of observers
in pockets of type $j$. We have shown that both of these options
would lead to inconsistencies.

With the analytic approximations we have suggested, the
distribution (\ref{smry}) can be readily calculated in a variety
of models. The probability distribution for the measurements of
observers in a given reference class can be obtained by
substituting (\ref{smry}) into Eq.~(\ref{calP}). Application to
examples which may be relevant to the landscape of string theory
(such as the Bousso-Polchinski scenario \cite{Bousso}) is left for
further research.

\section*{Acknowledgments}

J.G. is partially supported by grants FPA 2004-04582-C02-02 and DURSI
2001-SGR-0061. A.V. is partially supported by the NSF.

\section*{Note added}

After this paper was submitted, Easther,
Lim, and Martin (ELM) suggested an alternative prescription
for computing the weights $p_{j}$ assigned to different types of
pocket universes~\cite{ELM05}. In this note we show that their
prescription is equivalent to the CHC method.

The ELM prescription is formulated for a model with bubble nucleation,
and we shall restrict our attention to that case.
The main problem of counting bubbles is to select a large but finite
subset from the infinite set of all the bubbles created during the
infinitely long evolution. Once a finite subset is selected, the probability
ratio $p_{j}/p_{k}$ is found as the ratio of the number of bubbles
of types $j$ and $k$ from the subset, in the limit of large subset
size. The ELM proposal consists of choosing a large but finite number
of randomly drawn comoving worldlines and selecting the subset of
bubbles that intersect at least one of these worldlines. The distribution
of worldlines is assumed to be statistically independent of the bubble
nucleation process. Then the resulting weights $p_{j}$ are independent
of the choice of worldlines. 

The subset of bubbles selected by the ELM prescription differs from
the subset selected by the CHC method because some bubbles in the
ELM subset may have an arbitrarily small comoving size. However, we
shall now show that the ELM prescription produces the same result
for $p_{j}/p_{k}$ as the CHC method, for a generic bubble nucleation
model (with or without recycling).

The above mentioned assumption of statistical independence
(``the worldlines do not know about the pockets'') is equivalent to
assuming that there exists a well-defined probability density
$R(j;t)dt$ for a randomly chosen worldline from the congruence to end
in a bubble of type $j$ at a time $t$. Let us draw the
worldlines uniformly with a constant probability
per unit comoving 3-volume; this is an admissible distribution
in the sense of the ELM prescription. Then $R(j;t)$ is equal to the probability
of nucleating a bubble of type $j$ at time $t$ at a randomly chosen
comoving point. It is clear from Eq.~(\ref{dfdt}) that \begin{equation}
R(j;t)=\sum_\alpha \kappa_{j\alpha} f_\alpha (t), \label{Rjt ans}
\end{equation}
where the index $\alpha$ spans only the non-terminal vacua.

To simplify the calculation, we consider a uniform cubic grid of worldlines
where all the cubes have sides $\varepsilon/\sqrt{3}$. Then the smallest
and the largest (comoving) distances between neighboring worldlines
are $\varepsilon/\sqrt{3}$ and $\varepsilon$. This is again an admissible
congruence of worldlines, chosen independently of the bubble nucleation
process. The subset of bubbles selected by the ELM prescription will
include \emph{all} bubbles of comoving size $L\geq\varepsilon$ (the
counting of which constitutes the CHC prescription), as well as some
bubbles of smaller comoving size. So it remains to show that the additional
counting of bubbles of smaller size will not influence the CHC result,
in the limit $\varepsilon\rightarrow0$.

The total number of selected bubbles with sizes $L<\varepsilon$
is no greater than the 
%total
number of worldlines that do not
intersect larger bubbles of size $L>\varepsilon$. It follows from
Eqs.~(\ref{asympt}), (\ref{Rjt ans}), and the condition $f_\alpha^0=0$,
that the probability $R(j;t)$ of a worldline
encountering a bubble at late times $t$ is exponentially
small, $\propto e^{-qt}$.
Due to Eq.~(\ref{t cutoff}), the
limit $\varepsilon\rightarrow0$ corresponds to the limit
$t\rightarrow\infty$.  Hence, the additional counting of smaller-size
bubbles produces an exponentially small correction to the CHC
counting, and this correction disappears as
$t\rightarrow\infty$. Therefore, we have recovered the CHC bubble
counting result in the limit $\varepsilon\rightarrow0$.  Since the ELM
prescription is insensitive to the distribution of the worldlines, we
conclude that the two prescriptions always yield identical results.

\section*{Appendix A: FABI model with recycling}

We now consider a version of the $FABI$ model which includes
recycling between vacua $F$ and $I$ as indicated in the schematic $A
\leftarrow F \leftrightarrow I \rightarrow B$.

Writing out the master equation, we obtain: \beqa
\label{equn:evolve1} \frac{df_F}{dt}=
-(\kappa_{AF}+\kappa_{IF})f_F+\kappa_{FI}f_{I}
\\ \label{equn:evolve2} \frac{df_I}{dt}=
-(\kappa_{FI}+\kappa_{BI})f_I+\kappa_{IF}f_{F}
\\ \label{equn:evolve3}
\frac{df_A}{dt}= \kappa_{AF}f_{F} \\
\label{equn:evolve4} \frac{df_B}{dt}= \kappa_{BI}f_{I}\eeqa

We will first consider the comoving probabilities.   Noting that
we take $\kappa_{mn}$ to be independent of time, we can integrate
equations (\ref{equn:evolve1}) to (\ref{equn:evolve4}) from some
initial time $t=0$ to $t=\infty$ to obtain
\beqa \label{equn:qcfabievolve} -f_F(t=0)=
-(\kappa_{AF}+\kappa_{IF})\tilde{\psi}_F+\kappa_{FI}\tilde{\psi}_{I}, \\
-f_I(t=0)=
-(\kappa_{FI}+\kappa_{BI})\tilde{\psi}_I+\kappa_{IF}\tilde{\psi}_{F},
\\
f_A(t=\infty)= \kappa_{AF}\tilde{\psi}_{F}, \\
f_B(t=\infty)= \kappa_{BI}\tilde{\psi}_{I},\eeqa
where we have defined
\beq\tilde{\psi}_i\equiv\int_0^{\infty}f_i dt.\eeq

For simplicity we also define
 \beqa \label{equn:transratea}\kappa_{AF} \equiv a \\ \label{equn:transrateb} \kappa_{IF} \equiv
 b\\\label{equn:transratec}
\kappa_{FI} \equiv c\\ \label{equn:transrated} \kappa_{BI} \equiv
d \eeqa

Solving this system of linear equations for $\tilde{\psi}_i$ in
terms of the initial conditions and the transition rates, we find
the ratio of comoving probabilities as \beq \frac{p_A^c}{p_B^c} =
\frac{f_A(t=\infty)}{f_B(t=\infty)} = \frac{a(c+d f_F(t=0))}{d(b+a
f_I(t=0))},\eeq with $a,b,c,d$ related to the transition rates as
given in Eqs.~(\ref{equn:transratea})-(\ref{equn:transrated}).

For the initial condition $(f_F, f_I, f_A, f_B)(t=0)=(1,0,0,0)$, we have
\beq\frac{p_A^c}{p_B^c}=\frac{a(c+d)}{db}\eeq

If we consider $c \ll d$, we find
\beq\frac{p_A^c}{p_B^c}\approx\frac{a}{b}=\frac{\kappa_{AF}}{\kappa_{IF}}\eeq
in agreement with Eq.~(\ref{pApBFABI}).

We now compare this result with the CHC method. We solve for the
eigenvalues and eigenvectors of $\mathbf{M}$, yielding explicit
expressions for the $s_j$.
The result is \beqa
q = \frac{1}{2}(a+b+c+d-{\cal R}) \\
s_F = \frac{-q}{2bd}(a+b-c-d-{\cal R})\\
s_I = \frac{q}{d}\\
s_A = \frac{a}{2bd} (a+b-c-d-{\cal R})
\\ s_B = -1
\eeqa where \beq {\cal R}\equiv \sqrt{(a+b-c-d)^2+4bc}, \eeq $-q$
is the dominant eigenvalue, and $\textbf{s}=(s_F,s_I,s_A,s_B)$ is
the corresponding eigenvector. For completeness we note that the
other nonzero eigenvalue (call it $-q'$) is given by \beq q' =
\frac{1}{2}(a+b+c+d+{\cal R}) \eeq and that this differs from $q$
only by the sign of the square root term.

If we now want to calculate the probabilities in the limit that $c \ll
a,b,d$ (that is, we assume that recycling from $I$ back into $F$ is
nearly negligible), we can Taylor expand $s_A$ about $c=0$. We find
\beq
s_A \approx \frac{a}{2bd} (a+b-d-\sqrt{(a+b-d)^2}) -
\frac{ca}{d(a+b-d)}.\\ \eeq

Using Eq.~(\ref{pJaume}) we find as $c \rightarrow 0$, \beq
\frac{p_A}{p_B}\approx \left(\frac{H_F}{H_I}\right)^q \frac{ca}{d(a+b)} \to 0
\eeq if $a+b>d$ and \beq \frac{p_A}{p_B}\approx
\left(\frac{H_F}{H_I}\right)^q \frac{a}{bd}(-a-b+d) \eeq if
$a+b<d$, in agreement with the results of Section V, where $c$ was neglected from the very beginning.

\section*{Appendix B: Nondegeneracy of the subleading eigenvalue }

In this appendix we study the mathematical properties of the matrix
$\mathbf{M}$ defined by Eq.~(\ref{Mij}).
Formulations and proofs of some technical lemmas will be deferred
until the end of this appendix. We shall show that all nonzero eigenvalues
of $\mathbf{M}$ have negative real parts, and that the subleading
eigenvalue $-q$ of the matrix $\mathbf{M}$ is real and nondegenerate,
under the following assumptions:

I. The set of all the inflating (nonterminal) vacua cannot be split
into disconnected groups, where each vacuum from one group never nucleates
any vacua from other groups.

II. Transitions between any two inflating vacua are reversible: If
$\kappa_{mn}\neq0$, then $\kappa_{nm}\neq0$ as well.

III. There exist transitions to \emph{some} terminal vacua with nonzero
rates.

We begin by considering the vector of comoving volume fractions $\mathbf{f}$
which satisfies Eq.~(\ref{matrixf}),
\begin{equation}\frac{d}{dt}\mathbf{f}=\mathbf{Mf},\label{eq:master}
\end{equation}
where the matrix elements of $\mathbf{M}$ are defined
by\begin{equation}
M_{ij}=\kappa_{ij}-\delta_{ij}\sum_{r}\kappa_{ri}.\label{eq:Mij
wij rel}
\end{equation}
Conservation of the comoving volume requires that the sum of any column
of $\mathbf{M}$ be equal to zero. It follows from this condition
that all eigenvalues of $\mathbf{M}$ have nonpositive real parts
(see Lemma~1 below).

Suppose that there are $n_{i}$ inflating vacua and $n_{t}$ terminal
vacua. We can split the vector $\mathbf{f}$ into a direct sum of
two vectors, $\mathbf{f}_{(i)}$ and $\mathbf{f}_{(t)}$, representing
the volume fractions in these two types of vacua. Then Eq.~(\ref{eq:master})
can be rewritten as a system of two vector equations,
$$
\frac{d}{dt}\mathbf{f}_{(i)}  =\mathbf{R}\mathbf{f}_{(i)},\quad
\frac{d}{dt}\mathbf{f}_{(t)}  =\mathbf{S}\mathbf{f}_{(i)},
$$
where the matrix $\mathbf{R}$ describes the transition rates between
the inflating vacua and $\mathbf{S}$ describes the transition rates
from inflating to terminal vacua. Thus the matrix $\mathbf{M}$ has
the following block appearance,\begin{equation}
\mathbf{M}=\left(\begin{array}{cc}
\mathbf{R} & 0\\
\mathbf{S} & 0\end{array}\right).\end{equation}
Note that the matrix $\mathbf{R}$ is square with dimensions $n_{i}\times n_{i}$,
while the matrix $\mathbf{S}$ has dimensions $n_{i}\times n_{t}$
and may not be square.

We already know that the matrix \textbf{$\mathbf{M}$} has a zero eigenvalue and
no positive eigenvalues. We shall now prove, under the assumptions I-III
above, that the eigenvalue of  $\mathbf{M}$ with the algebraically largest
real part is a nondegenerate and negative eigenvalue $-q<0$.

It follows from the assumptions I and II that any inflating vacuum
will eventually nucleate bubbles with any other inflating vacuum (either
directly or after passing through bubbles of other inflating vacua).
The technical term for this property is that the matrix $\mathbf{R}$
is \textsl{irreducible}: for any indices $i,j$ there exist a chain
of indices $k_{1},k_{2},...,k_{s}$ such that all the matrix elements
$R_{ik_{1}},R_{k_{1}k_{2}},...,R_{k_{s}j}$ are nonzero.

Let us now analyze the eigenvalues of $\mathbf{M}$ in general. Any
eigenvector of \textbf{$\mathbf{M}$} has either all zero inflating
components, i.e.~it is a vector of the form $\mathbf{f=}\left(0,\mathbf{f}_{(t)}\right)$,
or $\mathbf{f=}\left(\mathbf{f}_{(i)},\mathbf{f}_{(t)}\right)$ with
$\mathbf{f}_{(i)}\neq0$. All vectors of the form $\left(0,\mathbf{f}_{(t)}\right)$
are obviously eigenvectors of $\mathbf{M}$ with eigenvalue $0$,
while eigenvectors of the form $\mathbf{f}=\left(\mathbf{f}_{(i)},\mathbf{f}_{(t)}\right)$
such that $\mathbf{Mf}=\lambda\mathbf{f}$ must satisfy $\mathbf{R}\mathbf{f}_{(i)}=\lambda\mathbf{f}_{(i)}$.
We shall shortly demonstrate that all eigenvalues of the matrix $\mathbf{R}$
have negative real parts and that the maximal eigenvalue $-q$
of $\mathbf{R}$ is real, negative, and nondegenerate, while the corresponding
eigenvector $\mathbf{r}$ can be chosen with all positive components.
It will then follow that the subleading eigenvalue of $\mathbf{M}$
is equal to $-q$ and is nondegenerate, and the corresponding
eigenvector of $\mathbf{M}$ is $\mathbf{s}=(\mathbf{r},-q^{-1}\mathbf{Sr})$.
Note that the components of the matrix $\mathbf{S}$ are nonnegative, and thus the inflating
components of the vector $\mathbf{s}$ have the opposite sign as compared
with the terminal components. This confirms our earlier statements~(\ref{monk}) and (\ref{monkey}).

Now we shall show that all eigenvalues of the matrix $\mathbf{R}$
have strictly negative real parts. Since $\mathbf{R}$ is irreducible
and has nonnegative off-diagonal components, there exists a nondegenerate
eigenvalue $\lambda_{0}$ of $\hat{R}$ with the largest real part
(the maximal eigenvalue) and all other eigenvalues have a smaller
real part (see Lemma~1). By construction, the matrix $\mathbf{R}$
is of the form~(\ref{eq:Mij wij rel}), where the indices $m,n$
are restricted to the range $1\leq(m,n)\leq n_{i}$. It follows from
the assumption III that the sums of each column of $\mathbf{R}$ are
not all equal to zero and thus \begin{equation}
\sum_{i=1}^{n_{i}}R_{ij}=-\sum_{i=n_{i}+1}^{n_{i}+n_{t}}\kappa_{ij}\leq0.\end{equation}
 Then it follows from the estimate~(\ref{eq:Min est}) of Theorem~1
that $\lambda_{0}\leq0$. However, we would like to prove the strict
inequality $\lambda_{0}<0$. To this end, we shall assume that $\lambda_{0}=0$
and arrive to a contradiction. If $\lambda_{0}=0$, there exists an
($n_{i}$-dimensional) eigenvector $\mathbf{v}$ with all positive
components (Lemma~1) such that $\mathbf{Rv}=0$. We can now express
the matrix $\mathbf{R}$ as a difference of a matrix $\mathbf{P}$
having vanishing sums of each column, and a diagonal matrix $\mathbf{Q}$
whose entries are nonnegative, namely\begin{align}
R_{ij} & =P_{ij}-Q_{ij},\\
P_{ij} & \equiv\kappa_{ij}-\delta_{ij}\sum_{l=1}^{n_{i}}\kappa_{li},\quad\sum_{i=1}^{n_{r}}P_{ij}=0,\\
Q_{ij} & \equiv\delta_{ij}\sum_{l=n_{i}+1}^{n_{i}+n_{t}}\kappa_{li}.\end{align}
Here the matrix $\mathbf{P}$ describes the nucleation of inflating
vacua and the matrix $\mathbf{Q}$ describes the nucleation of terminal
vacua. The equality $\mathbf{Rv}=0$ then yields \begin{equation}
\mathbf{Pv}=\mathbf{Qv}\geq0.\end{equation}
Since all the components of $\mathbf{v}$ are positive, and since
$Q_{ij}\geq0$ and $\mathbf{Q}\neq0$ by assumption III, we conclude
that not all components of the vector $\mathbf{Pv}$ are zero: \begin{equation}
\mathbf{Pv}\neq0.\label{eq:Pvneq0}\end{equation}
Note that the matrix $\mathbf{P}$ definitely has a zero eigenvalue since the sums of all columns vanish. Applying Lemma~1 to the matrix $\mathbf{P}$, we find that the maximal
eigenvalue of $\mathbf{P}$ is $\lambda_{P}\leq 0$, and since the matrix
$\mathbf{P}$ has a zero eigenvalue, this eigenvalue must be maximal, i.e.~$\lambda_P=0$. Intuitively, we expect
that this maximal eigenvalue is diminished when we subtract a nonnegative and nonzero matrix
$\mathbf{Q}$ from $\mathbf{P}$. Indeed, Lemma~3 shows that the
existence of a positive vector $\mathbf{v}$ satisfying $\mathbf{Pv}\geq0$
means that $\mathbf{v}$ is the eigenvector of $\mathbf{P}$ corresponding
to $\lambda_{P}=0$, in other words, we have $\mathbf{Pv}=0$, which
contradicts Eq.~(\ref{eq:Pvneq0}). Therefore the matrix $\mathbf{R}$
cannot have an eigenvalue $\lambda_{0}=0$.

Finally, we list some statements used in this appendix. A matrix (or
a vector) is called \textsl{nonnegative} if all components are nonnegative.
We write $\mathbf{a}\geq\mathbf{b}$ for vectors if $a_{i}\geq b_{i}$
for all $i$.

\textbf{Theorem 1: }
A nonnegative matrix $A_{ij}$ has a real eigenvalue $\lambda_{0}\geq0$
such that all other eigenvalues $\lambda_{i}$, $1\leq i\leq n-1$
are smaller than $\lambda_{0}$ in magnitude, $\left|\lambda_{i}\right|<\lambda_{0}$.
If the matrix $A_{ij}$ is irreducible, then the eigenvalue $\lambda_{0}$
is nondegenerate and the corresponding eigenvector can be chosen with
all positive components. Furthermore, if we denote \begin{equation}
\sigma_{j}\equiv\sum_{i}A_{ij},\label{Sigma}\end{equation}
 then the largest eigenvalue $\lambda_{0}$ is bounded by \begin{equation}
\min_{j}\sigma_{j}\leq\lambda_{0}\leq\max_{j}\sigma_{j}\text{.}\label{eq:Min est}\end{equation}

This is the Perron-Frobenius theorem and its corollary. For a proof,
see~\cite{Lanc69}, chapter 9, and also~\cite{Minc88}, chapter
1.

\textbf{Lemma 1:}
If a matrix $M_{ij}$ is irreducible and $M_{ij}\geq0$ for $i\neq j$,
then there exists a nondegenerate real eigenvalue $\lambda_{0}$ of
$M_{ij}$ such that the corresponding eigenvector has all positive
components, and all other eigenvalues have smaller real parts, $\textrm{Re}\,\lambda_{i}<\lambda_{0}$
(maximal eigenvalue). Moreover, if $\sum_{i}M_{ij}\leq0$ for all
$j$, then $\lambda_{0}\leq0$.

\textbf{Proof:} We can choose a real number $r$ such that the auxiliary
matrix $A_{ij}\equiv M_{ij}+r\delta_{ij}$ is nonnegative. The matrix
$A_{ij}$ is irreducible and thus, by Theorem~1, has an eigenvalue
$\tilde{\lambda}_{0}$ such that the corresponding eigenvector has
all positive components, and all other eigenvalues $\tilde{\lambda}_{i}$
lie within the circle $|\tilde{\lambda}_{i}|<\tilde{\lambda}_{0}$;
hence, $\textrm{Re}\,\tilde{\lambda}_{i}<\tilde{\lambda}_{0}$. Since
the eigenvalues of $M_{ij}$ are $\lambda_{i}=\tilde{\lambda}_{i}-r$,
the first statement follows. The second statement follows from the
estimate in Theorem~1, which yields\begin{equation}
\tilde{\lambda}_{0}\leq\max_{j}\sum_{i}A_{ij}=r+\max_{j}\sum_{i}M_{ij}\leq r,\end{equation}
hence $\lambda_{0}=\tilde{\lambda}_{0}-r\leq0$.

\textbf{Lemma 2: }
If $A_{ij}$ is a nonnegative irreducible matrix, and if $\lambda_{0}$
is the maximal eigenvalue of $A_{ij}$ from Lemma~1, and if there exists
a nonnegative vector $\mathbf{v}\neq0$ such that $\mathbf{Av}-\lambda_{0}\mathbf{v}\geq0$,
then $\mathbf{v}$ is an eigenvector of $\mathbf{A}$ with eigenvalue
$\lambda_{0}$.

A proof of this technical statement is contained in the proof of Theorem
4.1 in Ref.~\cite{Minc88}, chapter 1.

\textbf{Lemma 3:}
If a matrix $M_{ij}$ is irreducible and $M_{ij}\geq0$ for $i\neq j$,
and if $\lambda_{0}$ is the maximal eigenvalue of $M_{ij}$ from
Lemma~1, and if there exists a nonnegative vector $\mathbf{v}\neq0$
such that $\mathbf{Mv}\geq\lambda_{0}\mathbf{v}$, then $\mathbf{v}$
is an eigenvector with eigenvalue $\lambda_{0}$.

\textbf{Proof:} We consider the auxiliary nonnegative matrix $\mathbf{A}\equiv\mathbf{M}+r\mathbf{1}$
as in the proof of Lemma~1. It follows that $\mathbf{Av}-\left(\lambda_{0}+r\right)\mathbf{v}\geq0$.
We note that $\lambda_{0}+r$ is the maximal eigenvalue of $\mathbf{A}$.
Now Lemma~2 can be applied to the matrix $\mathbf{A}$ and it follows
that $\mathbf{v}\neq0$ is an eigenvector of $\hat{A}$ with eigenvalue
$\lambda_{0}+r$.


\begin{thebibliography}{99}

%\cite{Bousso:2004fc}
\bibitem{Bousso}
  R.~Bousso and J.~Polchinski,
  %``The string theory landscape,''
%decided to remove this reference:  Sci.\ Am.\  {\bf 291}, 60 (2004);
  %%CITATION = SCAMA,291,60;%%
  %``Quantization of four-form fluxes and dynamical neutralization of the
  %cosmological constant,''
  JHEP {\bf 0006}, 006 (2000)
  [arXiv:hep-th/0004134].
  %%CITATION = HEP-TH 0004134;%%

\bibitem{duff}
M.~J.~Duff, B.~E.~W.~Nilsson, and C.~N.~Pope,
  %``Kaluza-Klein Supergravity,''
  Phys.\ Rept.\  {\bf 130}, 1 (1986).
  %%CITATION = PRPLC,130,1;%%

%\cite{Susskind:2003kw}
\bibitem{Susskind}
  L.~Susskind,
  %``The anthropic landscape of string theory,''
  arXiv:hep-th/0302219.
  %%CITATION = HEP-TH 0302219;%%



%\cite{Vilenkin:1983xp}
\bibitem{AV83}
  A.~Vilenkin,
  %``Quantum Fluctuations In The New Inflationary Universe,''
  Nucl.\ Phys.\ B {\bf 226}, 527 (1983).
  %%CITATION = NUPHA,B226,527;%%
%\cite{Linde:1986fe}
\bibitem{Linde86}
  A.~D.~Linde,
  %``Eternally Existing Self-reproducing Inflationary Universe,''
%  Phys.\ Scripta {\bf T15}, 169 (1987);
  %%CITATION = PHSTB,T15,169;%%
%\cite{Linde:1986fd}
  %``Eternally Existing Self-reproducing Chaotic Inflationary Universe,''
  Phys.\ Lett.\ B {\bf 175}, 395 (1986).
  %%CITATION = PHLTA,B175,395;%%
%  A.~D.~Linde,
  %``Eternal Chaotic Inflation,''
%  Mod.\ Phys.\ Lett.\ A {\bf 1}, 81 (1986).
  %%CITATION = MPLAE,A1,81;%%

%\bibitem{bostrom} For a discussion of reference classes and their significance, see
%N. Bostrom, \emph{Anthropic Bias: Observation Selection Effects}, New
%York: Routledge (2002).

\bibitem{alex} A. Vilenkin, Phys. Rev. Lett. {\bf 74}, 846 (1995).

%=======
%>>>>>>> problands9-jaume.tex
%\cite{Garriga:1998px}
\bibitem{GTV98}
  J.~Garriga, T.~Tanaka, and A.~Vilenkin,
  %``The density parameter and the Anthropic Principle,''
  Phys.\ Rev.\ D {\bf 60}, 023501 (1999)
  [arXiv:astro-ph/9803268].
  %%CITATION = ASTRO-PH 9803268;%%

%<<<<<<< problands9-orig.tex
%\cite{Vilenkin:1998kr}
\bibitem{AV98}
  A.~Vilenkin,
  %``Unambiguous probabilities in an eternally inflating universe,''
  Phys.\ Rev.\ Lett.\  {\bf 81}, 5501 (1998)
  [arXiv:hep-th/9806185].
  %%CITATION = HEP-TH 9806185;%%


%\cite{Vanchurin:1999iv}
\bibitem{VVW}
  V.~Vanchurin, A.~Vilenkin, and S.~Winitzki,
  %``Predictability crisis in inflationary cosmology and its resolution,''
  Phys.\ Rev.\ D {\bf 61}, 083507 (2000)
  [arXiv:gr-qc/9905097].
  %%CITATION = GR-QC 9905097;%%


%\cite{Garriga:2001ri}
\bibitem{GV01}
  J.~Garriga and A.~Vilenkin,
  %``A prescription for probabilities in eternal inflation,''
  Phys.\ Rev.\ D {\bf 64}, 023507 (2001)
  [arXiv:gr-qc/0102090].
  %%CITATION = GR-QC 0102090;%%


\bibitem{LLM94}%\cite{Linde:1993xx}
  A.~D.~Linde, D.~A.~Linde, and A.~Mezhlumian,
  %``From the Big Bang theory to the theory of a stationary universe,''
  Phys.\ Rev.\ D {\bf 49}, 1783 (1994)
  [arXiv:gr-qc/9306035];
  %%CITATION = GR-QC 9306035;%%

%\cite{Linde:1994gy}
\bibitem{LLM+}
  A.~D.~Linde, D.~A.~Linde, and A.~Mezhlumian,
  %``Do we live in the center of the world?,''
  Phys.\ Lett.\ B {\bf 345}, 203 (1995)
  [arXiv:hep-th/9411111].
  %%CITATION = HEP-TH 9411111;%%
%\cite{Linde:1995uf}
  A.~D.~Linde and A.~Mezhlumian,
  %``On Regularization Scheme Dependence of Predictions in Inflationary
  %Cosmology,''
  Phys.\ Rev.\ D {\bf 53}, 4267 (1996)
  [arXiv:gr-qc/9511058];
  %%CITATION = GR-QC 9511058;%%
  J. Garcia-Bellido and A. D. Linde, Phys.~Rev.~D {\bf 51}, 429 (1995).


%\cite{Guth:2000ka}
\bibitem{Guth}
  A.~H.~Guth,
  %``Inflation and eternal inflation,''
  Phys.\ Rept.\  {\bf 333}, 555 (2000)
  [arXiv:astro-ph/0002156].
  %%CITATION = ASTRO-PH 0002156;%%
\bibitem{Tegmark}
  M.~Tegmark,
  %``What does inflation really predict?,''
  JCAP {\bf 0504}, 001 (2005)
  [arXiv:astro-ph/0410281].
  %%CITATION = ASTRO-PH 0410281;%%



\bibitem{VW}
  S.~Winitzki and A.~Vilenkin,
  %``Effective noise in stochastic description of inflation,''
  Phys.\ Rev.\ D {\bf 61}, 084008 (2000)
  [arXiv:gr-qc/9911029].
  %%CITATION = GR-QC 9911029;%%

\bibitem{Starobinsky} A. A. Starobinsky, in {\em
Current Topics in Field Theory, Quantum Gravity and Strings,} Lecture Notes in Physics, edited by H.J. de Vega and N. Sanchez (Springer, Heidelberg, 1986).

%\bibitem{li95} See e.g. %\cite{Linde:1995ck}
%  A.~D.~Linde,
  %``Quantum cosmology and the structure of inflationary universe,''
 % arXiv:gr-qc/9508019.
  %%CITATION = GR-QC 9508019;%%

%\cite{Vilenkin:1999kd}
\bibitem{AV99}
  A.~Vilenkin,
  %``On the factor ordering problem in stochastic inflation,''
  Phys.\ Rev.\ D {\bf 59}, 123506 (1999)
  [arXiv:gr-qc/9902007].
  %%CITATION = GR-QC 9902007;%%

\bibitem {GLM87}  A. S. Goncharov, A. D. Linde, and V. F.  Mukhanov,  Int. J.  Mod.~Phys.~A {\bf 2}, 561 (1987). 

\bibitem{Sasaki}
K. Nakao, Y. Nambu and M. Sasaki, Prog. Theor. Phys. {\bf 80}, 1041
(1988); Y. Nambu and M. Sasaki, Phys. Lett. {\bf B219}, 240 (1989).

\bibitem{open} %\cite{Bucher:1994gb}
J.~R.~Gott,
  %``Creation Of Open Universes From De Sitter Space,''
  Nature {\bf 295}, 304 (1982);
  %%CITATION = NATUA,295,304;%%
  M.~Bucher, A.~S.~Goldhaber, and N.~Turok,
  %``An open universe from inflation,''
  Phys.\ Rev.\ D {\bf 52}, 3314 (1995)
  [arXiv:hep-ph/9411206];
  %%CITATION = HEP-PH 9411206;%%
  K.~Yamamoto, M.~Sasaki, and T.~Tanaka,
  %``Large angle CMB anisotropy in an open university in the one bubble
  %inflationary scenario,''
  Astrophys.\ J.\  {\bf 455}, 412 (1995)
  [arXiv:astro-ph/9501109].
  %%CITATION = ASTRO-PH 9501109;%%

\bibitem{quasiopen} J.~Garcia-Bellido, J.~Garriga, and X.~Montes,
  %``Quasi-open inflation,''
  Phys.\ Rev.\ D {\bf 57}, 4669 (1998)
  [arXiv:hep-ph/9711214].
  %%CITATION = HEP-PH 9711214;%%

\bibitem{EWeinberg}
  K.~M.~Lee and E.~J.~Weinberg,
  %``Decay Of The True Vacuum In Curved Space-Time,''
  Phys.\ Rev.\ D {\bf 36}, 1088 (1987).
  %%CITATION = PHRVA,D36,1088;%%

\bibitem{recycling}
  J.~Garriga and A.~Vilenkin,
  %``Recycling universe,''
  Phys.\ Rev.\ D {\bf 57}, 2230 (1998)
  [arXiv:astro-ph/9707292].
  %%CITATION = ASTRO-PH 9707292;%%

\bibitem{BGV} A.~Borde, A.~H.~Guth, and A.~Vilenkin,
  %``Inflationary space-times are incomplete in past directions,''
  Phys.\ Rev.\ Lett.\  {\bf 90}, 151301 (2003)
  [arXiv:gr-qc/0110012].
  %%CITATION = GR-QC 0110012;%%

\bibitem{AV84}A.~Vilenkin,
  %``Quantum Creation Of Universes,''
Phys.\ Rev.\ D {\bf 30}, 509 (1984).
  %%CITATION = PHRVA,D30,509;%%

\bibitem{Linde84}
A. D. Linde, Lett.~Nuovo Cimento {\bf 39}, 401 (1984).

\bibitem{LLM2}
A.D. Linde, D.A. Linde and A. Mezhlumian, Phys. Lett. {\bf B345}, 203
(1995).

\bibitem{Lanc69}P. Lancaster, \emph{Theory of matrices}, Academic Press, NY, 1969.
\bibitem{Minc88}H. Minc, \emph{Nonnegative matrices}, Wiley, NY, 1988.

\bibitem{ELM05}R. Easther, E. A. Lim, and M. R. Martin, preprint astro-ph/0511233.


\end{thebibliography}
\end{document}